\documentstyle[prl,aps]{revtex}

\begin{document}
\renewcommand{\thefigure}{\arabic{figure}}
\def\reflabel#1{\noexpand\llap{\string\string\string#1\hskip.31in}}

\draft
\flushbottom
\twocolumn
[
\title{Avalanche Dynamics in Evolution, Growth, and Depinning Models}
\author{Maya Paczuski, Sergei Maslov*, and Per Bak}
\address{Department of Physics, Brookhaven National Laboratory \\
Upton, NY 11973 \\
** and Department of Physics, State University of New York at Stony
Brook, Stony Brook \\
New York 11794 \\
email: maya@cmt1.phy.bnl.gov maslov@cmth.phy.bnl.gov bak@cmth.phy.bnl.gov \\}
\maketitle
\tightenlines
\widetext
\advance\leftskip by 57pt
\advance\rightskip by 57pt
\begin{abstract}

The dynamics of complex systems in nature often occurs in terms of
punctuations, or avalanches, rather than following a smooth, gradual
path.
A comprehensive
theory of avalanche dynamics in models of growth,
interface depinning, and evolution is presented.
Specifically, we include the Bak-Sneppen evolution model,
the Sneppen interface depinning model, the Zaitsev flux creep model,
invasion percolation, and several other depinning models into
a unified treatment encompassing a large class of far from equilibrium
processes.  The formation of fractal structures, the appearance
of $1/f$ noise, diffusion with anomalous Hurst exponents, Levy
flights, and punctuated equilibria can all be related to the
same underlying avalanche dynamics.  This dynamics can be represented as
a fractal in $d$ spatial plus one temporal dimension.
 The complex state
can be reached either by tuning a parameter, or it can be
self-organized (SOC).  We present two {\it exact} equations for the
avalanche behavior in the latter case.
(1)  The slow approach to the critical attractor, i.e.
the process of self-organization, is governed
by a ``gap'' equation for the divergence of avalanche sizes.  (2) The
hierarchical structure of avalanches
is described by an  equation for the average number of sites covered
by an avalanche.
The exponent $\gamma$
governing the approach to the critical state appears
 as a constant rather than as
a critical exponent.  In addition,
the conservation of activity in the stationary state manifests itself
through the superuniversal result $\eta = 0$.
The exponent $\pi$ for the Levy flight
jumps between subsequent active sites can be
related to other critical exponents through a study of
``backward avalanches.''  We develop a scaling theory that
relates many of the
critical exponents in this broad category of extremal models,
representing different universality classes,
to two basic exponents characterizing
 the fractal attractor.
The exact equations and the derived set of scaling relations
are consistent with
numerical simulations of  the above mentioned models.

\end{abstract}
]
\narrowtext
\tightenlines


 \section{Introduction}

The term spatio-temporal complexity describes systems that contain
information over
a wide range of length and time scales
\cite{ zhang}.  Specifically,
if such a system
is studied through a magnifying glass, then no matter what the
level of magnification is, a variation of the image is seen as
different parts of the system are viewed.  Similarly, if the image of
a local part of the system is averaged over a time window, different
images are seen at different times, no matter how large
the time window is.  This contrasts with the
behavior of ordered systems, or
random systems, including chaotic ones, which become uniform when
viewed at sufficiently large length or time scales.

The  appearance of complexity in nature presents a fascinating,
long-standing puzzle.  In this article, a qualitative and quantitative
theory for the dynamics of complex systems is presented in the context
of simple mathematical models for biological evolution and growth
phenomena far from equilibrium.
Spatio-temporal complexity emerges as the
result of avalanche dynamics in driven systems.  We unify the origin
of fractals, $1/f$ noise, Hurst exponents for anomalous diffusion,
Levy flights, and punctuated equilibrium behavior, and elucidate
their relationships through analytical and numerical studies of
simple models, defined in Section II.  These phenomena are
signatures of spatio-temporal complexity and are related
via scaling relations to the fractal properties of the avalanches,
as summarized in Table \ref{ scalingrelations}.

We wish to focus on three general, empirical phenomena that
 are manifestations of complexity.
First, Mandelbrot \cite{ mandelbrot} has pointed out the
widespread occurrence of self-similar, fractal behavior in
nature.  Mountains, coastlines, and perhaps even the Universe \cite{ coleman}
have features at all scales.  River networks consist of streams of
all sizes \cite{ river} ; the pattern of earthquake faults, cracks in rocks,
 or oil reservoirs is
self-similar \cite{ barton}.  Second, noise with a $1/f^{\tilde d}$
power spectrum is emitted from a variety of sources,
including
light from quasars \cite{ press}, river flow \cite{ hurst}, and brain activity
\cite{ brain}.  Third,
many natural and social phenomena, including earthquakes, economic
activity, and biological evolution appear to evolve intermittently,
with bursts, or avalanches extending over a wide range of magnitudes,
rather than smoothly and gradually.  For
instance, the distribution of earthquake
magnitudes obeys the Gutenberg-Richter law \cite{ guten}.  Recent
experiments on slowly driven sandpiles \cite{ sandpile} and rice piles
\cite{ rice} show avalanches, or
disturbances in the heap, of all sizes.
Field {\it et al} monitored the dynamics of superconducting
vortices  and found the superconducting
analog of sandpile avalanches to be a power law over two
decades \cite{ field}.  Gould and Eldredge have proposed
that biological evolution
takes place in terms of punctuations, where many species become
extinct and new species emerge, interrupting quiet periods of
apparent equilibrium, known as stasis \cite{ gould}.  Large events, such as
massive extinctions or even the large scale structure
of the Universe, may have a statistical explanation
as the power-law tail of a Levy distribution.  This is the
case, even
though the large events are identifiable and tend to be viewed,
erroneously we believe, as ``atypical''.

 Even though spatio-temporal complexity is ubiquitous in nature, until
 recently, little understanding of its origin has been achieved.  One
 clear exception, though, are critical points for second order phase
 transitions in equilibrium systems, which are characterized by scale
 invariance.  Borrowing the terminology from equilibrium
 thermodynamics, we shall also refer to systems with a large range of
 length and time scales as ``critical.''  Spatial complexity occurs at
 the percolation transition in random bond or site models \cite{
 feder}; temporal complexity exists at the transition to chaos in the
 Feigenbaum map \cite{ chaos}.  These systems are critical, and thus
 complex, as the result of tuning: the temperature, or some other
 parameter, is set to a specific value in order to achieve
 criticality.  In nature, though, fine tuning of specific parameters
 is rare and unlikely; in addition, many systems in nature are far
 from an equilibrium state.

Thus another mechanism has been proposed;
 namely systems that are far from equilibrium become critical through
 self-organization \cite{ soc}.  They evolve through transient states,
 which are not critical, to a dynamical attractor poised at
 criticality.  In order for self-organization to occur, these systems
 must be driven slowly through a succession of metastable states.
The system jumps from one metastable state to another by avalanche
dynamics.  These avalanches build up long range
correlations in the system.
 Here, we shall be mainly concerned with the self-organized
 critical (SOC) origin of spatio-temporal complexity.  In
 some cases, though, similar considerations, such as scaling relations, can
 also be applied in cases where the criticality is tuned rather than
 self-organized.  In particular, this is relevant for systems which
undergo depinning transitions when a parameter is varied, such as interfaces,
charge density wave systems, and superconducting flux lattices.  In
this case, too, long lived metastable states are
important and the dynamics is that of avalanches \cite{ tangcdw}.
Avalanche dynamics in
these tuned depinning transitions are discussed in Section VI.

SOC complements the concept of chaos, where simple systems with few
degrees of freedom can display complicated behavior.  Chaos is associated
with a fractal ``strange'' attractor in phase space.  These self-similar
structures have little to do with fractals in large spatially extended
systems that have many degrees of freedom.  In addition, chaotic systems
exhibit white noise (e.g. a positive Lyaponov exponent)
 with limited temporal correlations whereas complex
systems have long range spatio-temporal correlations.

{}From the earliest BTW sandpile models \cite{ soc} and, later,
earthquake models \cite{ earthquake}, most of the evidence for SOC
behavior has been numerical.  Exceptions include the work on singular
diffusion by Carlson, Chayes, Grannan, and Swindle \cite{ singular},
and the one dimensional forest fire model \cite{ forest},
where exact results have been found by Drossel, Clar, and Schwabl
\cite{ dcs} and also by
Paczuski and Bak \cite{ pb93}.  Dhar \cite{ dhar} was able to
characterize the critical attractor of the BTW sandpile model in terms
of an Abelian group, and thereby calculate the number of states on the
attractor.  With Ramaswamy, he solved a directed sandpile model
exactly \cite{ dharrama}.  So far, though,
none of the solutions of these different
models have  yielded a general phenomenology for SOC behavior.
Most importantly, the fundamental mechanism for the
self-organization process, via avalanches, has not been well understood.
Pietronero and collaborators have developed a scheme,
the Fixed Scale Transformation \cite{ fst},
and applied it to a variety of nonequilibrium dynamical systems
including Diffusion Limited
Aggregation \cite{ dla},
 the sandpile models \cite{ Pietronerosand}, and the
Bak-Sneppen evolution model \cite{ Marsili}.   Here, we take a different
approach where, based on certain exact results together with a scaling
ansatz, we develop a theory that relates different critical exponents,
including the approach to the attractor, to two basic exponents which
are model dependent.

A common feature of many
 models exhibiting SOC is the selection of
extremal sites for the initiation of events, rather than statistically
typical sites.  This feature of extremal dynamics
 has been somewhat obscured by the
introduction of models, such as the BTW sandpile model or the forest
fire model, which appear to be driven randomly.   In the ``random''
BTW model, sand is added to random sites until a local threshold
is exceeded and a toppling occurs.
However, the statistics
of avalanches in the BTW model is not changed in
``deterministic'' versions where the
height of every site is raised uniformly until one site, the least
stable site, topples.   In the latter case, randomness only enters into
the initial conditions.   One might say that in the ``random'' BTW
model, the system selects the extremal sites itself, while in the
``deterministic'' case, it is
 forced to do so.
Similarly, in the Olami, Feder, Christensen
earthquake model \cite{ earthquake}, the force is raised
uniformly until the site with the largest force ruptures.  Zaitsev
\cite{ zaitsev}
introduced a model for low temperature flux creep where the motion
always takes place at the site with the largest force, and showed that
the model self-organizes to a critical state.  In the forest fire
model, it can be shown rigorously that the process is driven by the
burning of the largest forests with the oldest trees, despite the fact
that the trees grow randomly \cite{ dcs, pb93}.  Again, burning trees
randomly only serves as a vehicle for the system to burn the largests
forests.  In a remarkable paper, Miller {\it et al}
\cite{ McWhorter} suggested
through a different line of thinking that $1/f$ noise and fractality
are related to the preferential selection and thermal activation of
extremal sites.

Recently, a variety of simple models with
extremal dynamics that exhibit SOC have been introduced.
These models,
including the Bak-Sneppen evolution model
\cite{ bsmodel}, the Sneppen interface
depinning model \cite{ sneppen},  and the Zaitsev model
\cite{ zaitsev}, are defined in Section II.  They are
driven by sequentially
updating the site with globally extremal values of the signal.  This
information is propagated through the system via
 local interactions.  These models, representing different
universality classes,  are nevertheless similar to
invasion percolation\cite{ invasion,  lenormand}.
  Interestingly, invasion percolation was
recognized as an SOC phenomenon by Roux and Guyon \cite{ rouxguyon}
some years ago.  These authors, inspired by the sandpile analogy \cite{ soc},
defined avalanches in invasion percolation
in terms of a fluctuating extremal signal.  Some
additional new results for invasion
percolation have recently been presented by Maslov \cite{ maslov} using
methods similar to those applied in this work.

Our main analytical  results for this
class of models, including invasion percolation,
 are encapsulated by two exact equations plus a superuniversal
scaling law for stationary critical processes.
These results  address basic questions that arise
in SOC.  Perhaps the most obvious question
is how the system self-organizes into the complex, critical state.
This is described by a ``gap'' equation that relates the rate of approach
to the attractor to the average avalanche size
\cite{ gap}.  This equation
demonstrates that the stationary state of the system is a critical
state for the avalanches, where the average avalanche size diverges.
  Assuming that this divergence
has a characteristic exponent $\gamma$,
we show from the gap equation that the system
approaches its attractor algebraically with a characteristic exponent,
$\rho= 1/(\gamma -1)$,
for the transient.

The off-critical exponent
for the transient, $\gamma$,
can be expressed in terms of  the exponents in the stationary state
itself, via
a ``gamma'' equation for the hierarchy of
avalanches \cite{ gamma}.  This may
result from the fact that the off-critical direction is
stable (in SOC the critical point is an attractor for the
dynamics)  rather than being unstable.  Lastly, in SOC the critical
point is constrained to be stationary.  This leads to a fundamental
law for the critical states; $\eta =0$ in
all dimensions \cite{ eta}.
One dramatic consequence of this law is that the fractal dimension of
the active sites, $d_f$, is fixed by the probability distribution for
avalanche sizes  in the stationary state, i.e.
 $d_f = D(\tau -1)$.
 Thus the widespread appearance of fractal
structures can, perhaps, be viewed as a consequence of the existence of
a stationary limit.

By studying the space-time behavior of the activity
pattern in the critical state, i.e. the spatial location of the
extremal site at a particular point in time, one can describe
the activity pattern as a fractal embedded in $d$ spatial dimensions
plus one temporal dimension \cite{ mpb94}.   This fractal has a mass dimension,
or avalanche dimension $D$.   Long range time correlations, e.g.
$1/f$ noise, and spatial fractal behavior are unified as different
cuts in this underlying space-time fractal.  The temporal evolution
at a specific position is expressed as the activity along a one dimensional
time line piercing
the fractal perpendicular to the spatial dimensions.  The fractal spatial
structure  is found by cutting the fractal along
the spatial directions at that time.  We establish a formal
relation between $1/f$ type noise and fractal spatial behavior in terms
of the avalanche dimension $D$.  In the critical state,
the dynamics is best characterized
in terms of scale-free avalanches, initiated at extremal sites, and
propagating by an anomalous diffusion process.
 Fig. \ref{one} shows this space time fractal for the one dimensional
evolution model.

We have studied, in more detail, the value of the extremal
signal in time.
Time directed avalanches are naturally defined in
terms of this fluctuating signal \cite{ maslov}.  These avalanches
have a hierarchical structure of valleys within valleys.  Forward and
backward time directed avalanches have different statistical
distributions in the stationary state, reflecting the irreversibility
of extremal processes. The distribution of all forward avalanches, starting
at each update step for the extremal dynamics, is  a power law with
a superuniversal exponent, $\tau_f^{all} = 2$.
The distribution
 of all backward avalanches is also a power law, but with a different
model dependent exponent, $\tau_b^{all} = 3 - \tau$, where $\tau$ is
the usual avalanche size distribution exponent.

Taken together, these considerations lead to many scaling
relations for various physical quantities.  All of
the critical exponents that we consider
can be expressed in terms of two basic
exponents, for instance,
 the avalanche dimension $D$, and $\tau$, which characterizes
the distribution of avalanche sizes.  These scaling relations are
summarized in Table \ref{ scalingrelations}.     They
 provide numerous points to test
theoretical predictions with numerical simulations of different models.
We have made  numerical tests of
essentially all the scaling relations for many of the models
and find a pattern of consistency which confirms the
predicted scaling relations across different universality classes;
nevertheless more
accurate tests are needed for any specific result.
The results of our simulations are presented in Table \ref{ numerics},
and the error bars represent statistical errors for system sizes studied.
We urge that extensive, accurate simulations
be performed. Indeed, others  \cite{ grassberger}, \cite{ stanley}
 have already provided further
confirmation of our scaling relations.

 It is important
to note that for some models, the critical point can be
reached in different ways.
This is especially clear in the context of depinning.
The depinning transition can be self-organized
or reached by tuning either an external driving force or
the velocity.  In the context of interface
depinning and invasion percolation, the SOC version corresponds to
driving at constant velocity in the limit where the velocity
vanishes.
Some of the exponents are the same in the different cases
but others, in general, are different.  For example, $\eta =0$ for
all SOC depinning models, but may be non-zero for
the depinning transition at constant driving force.
The critical points that are reached by the self-organizing
process are different than the critical points that are reached in
an equivalent model by tuning a driving force.
Thus, despite the fact that these differences disappear in mean
field theory,  SOC cannot, even in principle,
 be viewed as sweeping a system
through a critical point, in  contrast to the claims in Ref. \cite{ sornette}.
The similarities and differences between
constant force and SOC depinning are elaborated in Section VI.
Our scaling relations are compared with recent numerical simulations
for constant force depinning \cite{ amaral}, \cite{ parallelstanley}.

Self-organized fractal growth
is fundamentally different than growth processes that are, for example,
described by (variants of) the Kardar-Parisi-Zhang (KPZ) equation
\cite{ kpz}, \cite{ generic}.
The KPZ equation is scale invariant by symmetry,
and thus the criticality
is not self-organized. It is essential to have long lived, metastable states
for the self-organization process to take place through
avalanches, without reverting to a
``ground state'' at or near equilibrium.
In addition, unlike
``generic scale
invariance'' \cite{ generic},
SOC does not require local conservation laws.  With the exception of
the Zaitsev model, all of the models we consider here are
nonconservative; in spite of this, they
can be shown to self-organize
to a critical state.  In the context of interface growth,
the dynamic scaling
approach  has been used to
separate  avalanche dynamics associated with
SOC and Langevin dynamics associated
generic scale invariance into distinct phenomenological
categories \cite{ maya}.

Here, we present a comprehensive and detailed account of our work on
the Sneppen interface model, the Bak-Sneppen evolution model,
invasion percolation, and
other SOC (and non SOC) critical models, including interface depinning.
  Some of these results have
been previously published in short accounts.  For clarity, we provide
here a complete, self-contained discussion of these models and our most
accurate and extensive numerical results.

Section II introduces the models for the general reader.
 Section III examines the
transient self-organization process and introduces the ``gap'' equation.
Section IV discusses the avalanche hierarchy
in the stationary state.  In particular we present the ``gamma'' equation,
the law for stationary states, $\eta=0$,
 and a discussion of time-directed avalanches.
The concept of backward avalanches is used to determine the exponent
$\pi$ governing the distribution of jumps in the activity, which has
a Levy distribution.
Section V unifies
spatial fractal behavior
with $1/f$ type noise.
Section VI contains our results on interface depinning.
 In the concluding section, the
new scaling relations that we derive
are summarized in Table
\ref{ scalingrelations}, and our
numerical results are summarized in Table \ref{ numerics}.
Appendix A explains in more detail the results for invasion percolation.

\section{ Definition of the Models}

In this section, we define all of the models studied here.

\begin{description}

\item {\bf Evolution \cite{ bsmodel}:} The Bak-Sneppen evolution model
 is perhaps the simplest
model of SOC.  In this `toy'
 evolution model, random numbers $f_i$ are assigned
independently to each site on a $d$-dimensional lattice of linear size
$L$. They are chosen from a uniform distribution between zero and one,
${\cal P}(f)$.  At each update step, the extremal site, i.e.  the one
with the smallest random number, $f_{min}$, is located.  That site,
and its $2d$ nearest neighbors are then assigned new random numbers,
drawn independently from the flat distribution
${\cal P}$.  After many updates have
occurred, the system reaches a statistically stationary state in which
the density of random numbers in the system vanishes for $f < f_c$ and
is uniform above $f_c$.  In the thermodynamic $L \rightarrow \infty$
limit, no random number with $f > f_c$ is ever the extremal site.  A
snapshot of the stationary state in a finite one dimensional system is
shown in Fig. \ref{two}.  Except for a localized region, the avalanche, where
there are small random numbers, all the random numbers in the system
have values above the self-organized threshold $f_c= 0.66702 \pm
0.00003$ in one dimension.

The evolution model exhibits punctuated equilibrium behavior as
does real biology \cite{ gould}.
It simulates
evolutionary activity  in terms of mutations of individual
species and interdependencies in a food chain.  The
random numbers represent the fitness of a species, and chosing
the smallest random number mimicks the ``Darwinian'' principle that the
least fit species is replaced or
 mutates.  The dynamical impact of the mutation of
the least fit species on
the rest of the ecology is simulated by changing the fitness of neighboring
species on the lattice.    Discussions of its possible connection to
biological evolution and macroeconomics may be found in Refs.
\cite{ bsmodel}, \cite{ biology}, and a mean field analysis in
\cite{ flyvbjerg}, \cite{ rayjan}, \cite{ stonybrook}.  A generalized
$M$ -vector
model where each species has many $(M)$ internal
degrees of freedom has very recently
been introduced and solved exactly \cite{ boettcher} in the
$M \rightarrow \infty$ limit.

\item {\bf Sneppen \cite{ sneppen}:}
In the Sneppen interface model for SOC depinning,
an interface of size $L^d$ defined on a discrete lattice $(\vec
x , h)$ moves under the influence of quenched random pinning forces $f(\vec x ,
h)$ assigned independently from the flat distribution
${\cal P}$.  Initially, $h(\vec x)=0$.
Growth occurs by advancing the extremal site on the interface with the
smallest random pinning force, $f_{min}$, by one step.  Then a
constraint is imposed for all nearest neighbor gradients, $|h(\vec x)
- h(\vec x')| \le 1$.  This is met by advancing the heights of
neighboring sites.  The process is repeated indefinitely.   Like
the evolution model, the Sneppen model also reaches a statistically stationary
state where the density of random pinning forces on the interface
vanishes for $f < f_c$ and is uniform above $f_c$.
In this state, the interface moves in bursts of localized activity,
as indicated schematically in Fig. \ref{schematic}.
Outside of these regions of activity, the interface is frozen for
potentially long periods of time until a burst moves into the frozen region.
The interface moves in a jerky, irregular manner,
rather than smoothly and gradually advancing as a whole.

Tang and Leschhorn
\cite{ sneppentl}
showed that the one-dimensional model, in the stationary state, identifies
from time to time with paths on a critical directed percolation cluster.
It has been proposed
that this identification with blocking paths can be used to obtain all of
the critical exponents in terms of two independent exponents \cite{ mppre}.
The identification with ``blocking surfaces''
is analogous to invasion percolation, where the invading region
identifies from time to time with a critical cluster of ordinary percolation
\cite{ invasion}, \cite{ rouxguyon}.  Possible physical realizations of
the Sneppen model and other interface depinning models are discussed in
Section VI.

\item {\bf Zaitsev \cite{ zaitsev}:}  Zaitsev introduced an
extremal  model for flux creep.
Random numbers $f_i$, chosen from $\cal P$, are assigned independently
to each site on a $d$ dimensional lattice of linear
size $L$.  At each update step,
the site with the
largest number is chosen, and a random number chosen from ${\cal P}$
is subtracted from that site and
equally distributed to  the $2d$ nearest neighbors.  The activity
conserves the sum of all the random numbers in the system.
 It has been suggested \cite{ roux} that this model is in the
same universality class as a self-organized
 ``linear'' interface depinning model (LIM), sometimes referred to
as the quenched Edwards-Wilkins \cite{ edwards} equation.

\item {\bf LIM:}  In the linear interface depinning model, the force at
each site has a random contribution, $f(\vec x,h)$, chosen from $\cal
P$, to represent quenched random pinning forces,
plus a linear configurational term $f_{conf} \sim \nabla^2 h$,
where $h$ is the local height and $\nabla^2$ represents a discretized
Laplacian.   Thus the local internal force is
\begin{equation}
F_{int}({\vec x},t)= \nabla^2 h({\vec x}, t) + f({\vec x},h({\vec x},t))
\qquad .
\label{e.force}
\end{equation}
We consider cases where this model
is  driven either (a) with extremal dynamics
\cite{ mpb94, roux}
 or (b) in a parallel non self-organized mode \cite{
leschhorn}.  In (a) the site with the maximum total force is advanced
by one unit, leading to  SOC, or constant velocity
 depinning.  In (b) the model may be tuned to
a depinning transition by adding an external driving force $F$ to all sites
and advancing in parallel every site, where the total force
$(F_{int} + F)$  is
positive, by one step.  When $F > F_c$ the
interface moves with a finite velocity,
while for $F < F_c$ it becomes stuck in a metastable
state.  At $F=F_c$ it undergoes a depinning transition.  A tuned depinning
transition may also be realized by externally setting the velocity,
$v$, of
the interface and allowing the force $F$ to fluctuate so as to maintain
that velocity.  The stationary state of the SOC version  corresponds to
the
depinning transition at constant velocity.  The SOC version tunes itself
to a constant velocity depinning transition.
In Section VI, a comparison is made between the LIM
and models for fluid invasion in porous media \cite{ porous}, \cite{ martys}
and interface depinning
in the Random Field Ising Model \cite{ rfim}.
There, it is also argued that
$\tau$ and $D$ are the same for the (constant force)
tuned and SOC variants, but
$\eta$ and other dynamical exponents (e.g. $d_f$ and $z$) can be different.

 \item {\bf TLB:} A  model for depinning
of an elastic interface at constant force that was studied
by Tang-Leschhorn \cite{ paralleltl} and also by Buldyrev {\it et al}
\cite{ parallelstanley}.
Again, an interface of size $L^d$ defined on a discrete lattice
$(\vec x , h)$ moves under the influence of random pinning forces
$f(\vec x , h)$ assigned independently from ${\cal P}$.  Initially, $h(\vec
x)=0$.
 Instead of
advancing the most unstable site, as in the Sneppen model, all
unstable sites with $f <F$ are advanced in parallel.  Then the
constraint is imposed for all nearest neighbor gradients, $|h(\vec x)
- h(\vec x')| \le 1$.  The system
is relaxed to meet the gradient constraint.   When $F=F_c$ the interface
undergoes a depinning transition. Tang and Leschhorn
\cite{ sneppentl}, \cite{ paralleltl} showed that
in the critical state
both the TLB model and the Sneppen model, in one dimension, identify
with a directed percolation cluster of sites with $f > f_c$.
In Section VI, it is argued that the
 exponents $\tau$ and $D$ are the same for the TLB and Sneppen
models; while $\eta$, $z$, and other dynamical exponents can
be different.

\item {\bf DP:} In directed percolation,
a preferred direction,
labeled by $t$, is chosen and bonds  are oriented with respect to $t$.
Percolation is only allowed in the direction of increasing $t$.
Each bond exists with probability $f$.  When $f=f_c$, the DP
cluster can become infinitely large.  This model can be
viewed as  the ``parallel'' constant force
version of the evolution
 model.
In DP, there is a critical point which
is tuned.

\item {\bf Invasion Percolation \cite{ invasion}, \cite{ lenormand}:}

Invasion percolation is the oldest member of this class of models.
It was studied as an SOC
phenomenon, in terms of avalanches, by Roux and Guyon \cite{ rouxguyon}.
In invasion percolation, random numbers $f_i$ are assigned
independently to each site on a $d+1$-dimensional lattice of linear
size $L$. They are
chosen from a uniform distribution between zero and one, ${\cal
P}(f)$.  Initially, one $d$-dimensional
side of the lattice is the ``invaded cluster''.
The random numbers at the boundary of the invaded cluster are
examined. At each update step, the extremal site, i.e.  the site with
the smallest random number, $f_{min}$, on the boundary of the
invaded cluster is located.  That site is then invaded, and new sites
susceptible to growth are added to the boundary.
The universality class of the model is sensitive to the exact
definition of the boundary.
Different definitions of the boundary  (IP-1, IP-2, IP-3)
are given in Appendix A, along with a discussion of their
physical realizations.

\end{description}


\section{Self-Organization}

Self-organization, as used here, refers to a dynamical process
whereby a system starts in a state with uncorrelated behavior and
ends up in a complex state with a high degree of correlation.
The time that it takes to self-organize grows as the system size
increases; so that for large systems self-organization is a slow
process.
In contrast to the earlier models of SOC, the process of self-organization
in the
extremal models that are discussed here is very well understood.
The stationary state is critical and is approached algebraically, through a
sequence of transient states.
In order to demonstrate this, we shall first
discuss self-organization in the context of the evolution model, which
has been defined in Section II, and then generalize the results to
other models.

Let us consider the situation where the distribution
of $f$'s initially is uniform in the interval $[0,1]$ in a $d$-dimensional
evolution model of linear size $L$.   Initially, the activity, i.e. the
spatial location of the smallest random number in time, jumps randomly
throughout the system.
Eventually, after a long transient, the system
reaches a complex state where the activity is
correlated, as shown in Fig. \ref{one}.
In order to study this transient process, we examine the
value of the minimal random number chosen, $f_{min}$, as a function
of sequential time, $s$, or the total number of updates.
This signal $f_{min}(s)$ can be related to the distribution of
random numbers in the system.

The first value, $f_{min}(0)$, to be chosen for
updating is ${\cal O}(L^{-d})$.   Since, by definition, there are no
random numbers in the system smaller than $f_{min}(0)$, this quantity
is defined to
be the initial value of the gap, $G$,
in the distribution of $f$'s; that is $G(0)= f_{min}(0)$.
Eventually, after $s$ updates, a larger gap
$G(s)> G(0)$ opens up.   The density of sites
with random numbers below $G$ vanishes in the thermodynamic
$L \rightarrow \infty $ limit, and is uniform above $G$.  The current
gap, $G(s)$,  is the maximum of all the
minimum random numbers chosen, $f_{min}(s')$, for all $0 \le s' \le
s$. Fig. \ref{three} shows $f_{min}$  as a function of $s$ during
the transient for a small system. The solid line shows
the gap, $G(s)$, as a step-wise
increasing function of $s$.   The gap is an envelope function that tracks
the increasing peaks in $f_{min}$.
Clearly, when the gap jumps to a new higher value, there are no
sites in the system with random numbers less than the gap.  Since the
distribution, $P(f)$,
 that each of the random numbers are chosen from is flat,
all of
the random numbers in the system are
 uniformly distributed in the
interval $[G(s),1]$ at those moments in time when the gap jumps.
 Thus, the
envelope function tracks the distribution of random numbers in
the system.
At the moments in time when the gap jumps, each of
the random numbers in the system is independently distributed in
the interval $[G(s),1]$.

By definition, the separate instances when
the gap $G(s)$ jumps to its next higher value are separated
by avalanches.
Avalanches correspond to plateaus in $G$ during which
$f_{min}(s) < G (s)$.  This guarantees that  the events
within a single avalanche are causally and spatially connected.
A new avalanche is initiated every time the gap jumps, and all the
consecutive
random numbers which are smaller than the gap after this
event must have originated from
the site that started the new avalanche.  Once an avalanche
is over it does not affect the behavior of any subsequent avalanche,
except in terms of the gap threshold.  As the gap increases,
the probability for the new random numbers to fall below the
gap increases also, and
longer and longer avalanches typically occur.

Since the distribution of random numbers above the gap is flat,
the average size of the jump in the gap at the
completion of each avalanche  is $(1-G(s))/L^d$.  The other
quantity that is needed in order to describe the self-organizing
system is the
average size of the plateau for a given value of $G$, or the
average avalanche size $<S>_{G(s)}$.   We shall prove
below that in the limit of large system
sizes $L$,
the
growth of the gap versus time, $s$, obeys  the
exact  gap
equation \cite{ gap}:
\begin{equation}
{d G(s) \over d s } =
 {1- G(s) \over L^d <S>_{G(s)}} \qquad .
\label{e.approach}
\end{equation}

As the gap increases, so does the average avalanche size
$<S>_{G(s)}$, which
eventually diverges as $G(s) \rightarrow f_c$.  Whereupon, the model
is critical and the process achieves stationarity. In the limit
$L \rightarrow \infty$, the density of sites with $f<f_c$ vanishes, and the
distribution of $f$'s is uniform above $f_c$.
The gap equation (\ref{e.approach}) defines the mechanism of
approach to the self-organized critical attractor.  It contains
the essential physics of SOC phenomena.  When the average avalanche
size diverges, $<S>_{G(s)} \rightarrow \infty$, the system becomes critical.
At the same time, $d G \over d s$ approaches zero,
which means that the system becomes stationary.
Examining Eq. (\ref{e.approach}) one notices another
mathematical possibility for the stationary limit, which does not require
diverging avalanches.  The time derivative
$d G \over d s$ is exactly zero if the current
gap $G(s)=1$. This situation, however, is not realised in
any extremal model in the limit that
 $L \rightarrow \infty$, since the presence of any interaction
between sites (such as replacing nearest neighbors with
new random numbers in the evolution model) will result in $f_c<1$.

We  prove the gap equation as follows:  for any selected resolution
$\Delta G \ll 1$ along the gap axis there is  a system size $L$
sufficiently large that many avalanches are needed to increase the
gap from $G(s)$ to $G(s)+ \Delta G$. In this case, the  sum of (a) the
temporal durations $S$ of an individual avalanche
and (b) the jumps in the gap at the end
of each avalanche will both average within this interval because they
are both sums of independent random variables.
Therefore, in the limit $L \rightarrow \infty$ Eq. (\ref{e.approach})
is exact.   To be more specific, suppose that the current value of the gap
in the system is $G(s)$. The average number of avalanches
required to increase the gap by $\Delta G$ is $N_{av}
=\Delta G L^d/(1-G(s))$.  By selecting
system size $L \gg \Delta G^{-1/d}$ we ensure that
$N_{av} \gg 1$.  $N_{av}$ can be made arbitrarily large by taking
the large $L$ limit.  In this limit,
the average number of time steps required to increase
the gap from $G(s)$ to $G(s)+ \Delta G$ is given by the interval
$\Delta s = \langle S \rangle_{G(s)}N_{av} =
\langle S \rangle_{G(s)} \Delta G L^d/(1-G(s))$, and from the law
of large numbers
the fluctuations of this interval around its average value
vanish.   This equation shows that
the ratio of the interval length $\Delta s$ to the time $s$ required
to reach $G(s)$, $\Delta s /s$, vanishes as $\Delta G \rightarrow 0$.
Rewriting the interval equation as
${\Delta G \over \Delta s } =
 {1- G(s) \over L^d <S>_{G(s)}}$ and taking the continuum limit
we restore the differential equation
(\ref{e.approach}).

In order to solve the gap equation we need to
know precisely how the average avalanche size $<S>_{G(s)}$ diverges
as the critical state is approached.  It is natural to assume that
this divergence has a power law form.  In analogy with percolation
clusters,  we now make an ansatz  that the average avalanche size
diverges  as $G$ approaches $f_c$
as
\begin{equation}
<S> \sim  (f_c-G)^{-\gamma} \quad .
\label{e.gammaa}
\end{equation}

We will show in the next section that for the extremal models studied
here the exponent $\gamma \geq 1$.  Let us first consider the
case $\gamma >1$.
Inserting Eq.\ (\ref{e.gammaa}) into
Eq.\ (\ref{e.approach}) and
integrating, one obtains
\begin{equation}
 \Delta f = f_c - G(s)
\sim ({s \over L^d})^{-\rho}\, , \quad {\rm where} \quad \rho =
 {1 \over { \gamma -1}} \; .
\label{e.trans}
\end{equation}
Eq. (\ref{e.trans})
 was also found by Ray and Jan \cite{ rayjan}.  It shows
that the critical point $(\Delta f = 0)$
is approached algebraically with
an exponent $-{1 \over \gamma-1}$.  Eq. (\ref{e.trans}) holds
over the range $L^d \ll s \ll L^{\tilde D}$.  The lower  limit
requires that the avalanches are in the scaling regime, so that
Eq. (\ref{e.gammaa}) is valid.  The upper limit requires that the
cutoff, $r_{co} \sim \Delta f ^{-\nu}$, in the spatial
extent of the avalanches is much less than the system size, or
$\Delta f ^{-\nu} \ll  L$.  Inserting Eq. (\ref{e.trans}) into
this expression gives
\begin{equation}
{\tilde D } = d + {\gamma - 1 \over \nu }\qquad .
\label{e.tildeD}
\end{equation}

The boundary case $\gamma =1$, realized for instance in the mean field
version of the evolution model
\cite{ flyvbjerg, rayjan, stonybrook}, results in an exponential relaxation
to the critical attractor:
\begin{equation}
\Delta f \sim e^{-As/L^d} \qquad ,
\label{e.trans2}
\end{equation}
where $A$ is a numerical constant that is independent of $L$.
This expression is valid as long as
$L^d \ll s \ll L^d\ln L$, and the system reaches the stationary state
when $s \sim L^d \ln L$.

It is straightforward to demonstrate, in
a step by step fashion, that the gap equation (\ref{e.approach}) holds
not only for the self-organization process in the evolution model, but also
for the Sneppen interface  model \cite{ mppre}.
For the self-organized LIM and the Zaitsev model,
the distribution
of internal forces in the system is not given by a flat distribution,
unlike the evolution and Sneppen models.  However, the extremal dynamics
allows one to define a fluctuating signal $f_{min}(s)$ and therefore
a gap function $G(s)$ as above.  Now, though, the average size in
the jumps of the gap is not given simply by $(1-G(s))/L^d$.  However,
as long as the distribution of internal forces above the gap is not
singular, when the avalanche is finished,
then a weaker form of the gap equation holds,
\begin{equation}
{d G(s) \over d s } \sim
 {1\over L^d <S>_{G(s)}} \qquad ,
\label{e2.approach}
\end{equation}
If the scaling ansatz Eq. (\ref{e.gammaa}) holds,
 then depending on the value
of $\gamma$, either Eq. (\ref{e.trans}) or Eq. (\ref{e.trans2}) describe the
approach to the critical state.

For $d+1$ dimensional
invasion percolation, the situation is slightly more complicated.
During the transient process, the size of the boundary
where growth may occur is growing.
It can be shown \cite { rouxguyon} that during the transient period
of invasion percolation starting from a flat
configuration, the active boundary $b(s)$
of the invading
cluster scales with the invaded volume $s$ as
\begin{equation}
b(s)/L^d \sim
(s/L^d)^g=(s/L^d)^{d_B-d \over D-d} \qquad .
\label{e.boundary}
\end{equation}
Here, $d_B$ is the fractal dimension of the boundary where growth
may occur and $D$ is the fractal dimension of the invaded cluster.
The derivation of this equation is explained in Appendix A.
The system reaches a stationary state
when $s \sim L^D$.  At this point, the size of the active boundary
also reaches
its finite size limit $L^{d_B}$.
The proper gap equation for  invasion percolation,
taking into account the growth of the active boundary with time,
can be written as
\begin{equation}
{d G(s) \over d s} \sim {1 \over L^d(s/L^d)^g<S>_{G(s)}}\qquad .
\label{e3.approach}
\end{equation}
In the  asymptotic critical region, the scaling ansatz
Eq. (\ref{e.gammaa}) can be inserted into the gap equation
(\ref{e3.approach}).
Integration
 gives
\begin{equation}
\Delta f=f_c-G(s)=({s \over L^d})^{-{1-g \over \gamma-1}} \qquad ,
\end{equation}
which holds for
$L^d \ll s \ll L^{d+{\gamma-1 \over \nu(1-g)}}$.

The scaling relations between these and other exponents are explained
in the next section.  One might naively suspect that
 the critical exponent $\gamma$ would be an independent
exponent describing off-critical behavior.
This is not the case. The results of the next
section allow us to determine $\gamma$ in terms of the avalanche
dimension $D$, and  the avalanche distribution exponent, $\tau$,
and solve the gap equation.


\section{Properties of the Stationary, Critical  State}

In all previous SOC models, like
for instance the BTW sandpile model
\cite{ soc} or the earthquake models \cite{ earthquake},
Self-Organized Criticality manifested itself in a power law
distribution of bursts of activity, or avalanches, following
a single perturbation.  The new series of extremal SOC models
is not an exception. In fact, avalanches in these models have
an additional hierarchical structure of subavalanches within
avalanches.  In this section, avalanches in the stationary state
are defined and related to the avalanches in  the transient
process defined in the previous section.  This enables us to
determine the exponent $\gamma$, characterizing the transient,
in terms of the stationary probability distribution of avalanches.
The hierarchical avalanche structure is revealed in the ``gamma''
equation \cite{ gamma}.  We then demonstrate the  law
for stationary states, $\eta =0$
\cite{ eta}.
This law implies, among other things, that the fractal dimension
of active sites is also determined by the stationary probability
distribution for avalanches through the scaling relation
$d_f = D(\tau -1)$.  All of the above mentioned results hold for
forward avalanches; we close this section with a discussion of
backward avalanches \cite{ maslov}, and a derivation of a scaling
relation for the Levy flight exponent, $\pi = 1 + D(2-\tau)$.

Avalanches in the stationary state
can be defined for any of the extremal models as follows:
Suppose that at time $s$ the smallest random number
in the system
was larger than $f_o$, where $0<f_o<1$ is an
auxiliary parameter used to define  avalanches.
Each of the new random numbers introduced at this time step
can be smaller than $f_o$ with probability
$f_o$ (for the flat distribution ${\cal P}(f)$).
According to the rules of the model, if one (or more) of the new
random numbers  is less than $f_o$, the smallest of those
will be selected at the next time step $(s+1)$.
This initiates a creation-annihilation branching process, where
the sites with $f_i<f_o$ play the role of particles, and the particle
with the smallest random number is chosen for
updating at each time step.
While the avalanche
 continues to run, there is at least one such ``particle''
and the global, minimal random number
is smaller than $f_o$. The avalanche
stops, say at time $S+s$,
 when the last ``particle'' with a random number smaller
than $f_o$ is eliminated.  By definition, the global, minimal
random number at this time step will be
larger than $f_o$ once again.  Thus, one can view the parameter
$f_o$ as a branching probability for the creation of particles.
We will call the avalanches, defined by this rule, $f_o$-avalanches.
They can be easily extracted from the
{\it temporal signal} of the model $f_{min}(s)$, which is the value
of the selected
minimal number as a function of time $s$. In terms of this signal, the
size of an $f_o$-avalanche
is  the number of time steps elapsed between subsequent punctuations
of the barrier $f_o$ by the signal $f_{min}$.  For the example given
here, the size of the avalanche is $S$.  The hierarchical structure
of  $f_o$- avalanches is demonstrated in Fig. \ref{five}.

The statistics of the avalanches clearly depends on the value of
the branching probability $f_o$. The larger it is, the larger
is the expected size of the avalanche.
In analogy with ordinary percolation \cite{ grimmett},
there must be a critical value $f_c <1$ of the branching probability
such that the expected size of the $f_c$- avalanche cluster becomes infinite.
That means that in the thermodynamic limit $L \rightarrow \infty$
in the stationary state of the system, with probability one, at least
one
"particle" with $f_i<f_c$ will exist, and the signal $f_{min}$
 will be smaller than $f_c$, also with probability one.
In analogy with ordinary percolation \cite{ grimmett}
and other ``branching
processes'' such as directed percolation
\cite{ feder}, a variety of nonequilibrium lattice models
\cite{ jensen},
 or the propagation of an
avalanche in the BTW sandpile model \cite{ soc}, we use
the following scaling ansatz for the probability
distribution $P(S,f_o)$ of
$f_o$- avalanches of size $S$:
\begin{equation}
 P(S,f_o)=S^{-\tau}g(S(f_c-f_o)^{1/\sigma})
\label{e.av.dist}
\end{equation}
Here, $\tau$ and $\sigma$ are model dependent exponents and
$g(x)$ is scaling function, which decays rapidly to zero  when
$x \gg 1$ and goes to a constant
when $x \rightarrow 0$.
This scaling ansatz for various models
has been confirmed by  numerical simulations
in \cite{ bsmodel},
\cite{ rayjan},  \cite{ stonybrook},
\cite{ sneppentl}, \cite{ martys}, \cite{ rfim}.
When the auxiliary parameter $f_o$ is selected to be equal
to $f_c$, the avalanche distribution  is a
power law $P(S) \sim S^{-\tau}$ without cutoff. When
the parameter $f_o$ is lowered below $f_c$ these critical avalanches
are subdivided into  smaller ones
and acquire a cutoff $S_{co}=(f_c-f_o)^{-1/\sigma}$.
The average size of an
 $f_o$-avalanche diverges as $f_o$ approaches $f_c$
as
\begin{equation}
<S> \sim  (f_c-f_o)^{-\gamma} \qquad .
\label{e.gamma}
\end{equation}
Simple integration of the power law (\ref{e.av.dist}) gives
the usual expression for $\gamma$ in terms
of $\tau$ and $\sigma$, as occurs for example in ordinary
percolation \cite{ feder}, \cite{ grimmett}:
\begin{equation}
 <S> = \int S  P(S,f_o)dS \qquad ; \qquad \gamma={2-\tau \over \sigma}
\qquad .
\label{e.gamma2}
\end{equation}

\subsection{ The BS Branching Process.}

Unlike the other extremal models, the
evolution model has an additional feature greatly simplifying
numerical studies of $f_o$- avalanches.
The propagation of an $f_o$- avalanche in the evolution
 model is totally independent
of the environment (the values of $f_i > f_o$), where it propagates.
This means that, with respect to an $f_o$- avalanche, all of the
sites with $f_i > f_o$ can be treated as vacuum sites.  Sites are unimportant
as long as they are not occupied with the ``particles'' of the $f_o$-
avalanche.
In order to study $f_o$- avalanches we have to keep
track only of
sites that have random numbers $f_i<f_o$.
In simulations of this BS branching process \cite{ gap},
which is {\it exactly} equivalent to an $f_o$- avalanche
in the evolution model, we first create $2d+1$ random numbers,
chosen from the flat distribution $\cal P$,
at the center of the lattice and its $2d$
nearest neighbors.  Random numbers smaller than the parameter
of the simulation,
$f_o$, are stored along with their positions. At each time step,
the minimal of the stored random numbers is ``activated'' until
there are no more
stored
random numbers.  At this point the $f_o$- avalanche is finished.
This method gives an efficient way to study the $f_o$- avalanche distribution
and other properties, completely free from system size corrections.
An avalanche is considered infinite, and not included in the distribution,
if its size is larger than a numerically imposed cutoff $s_{max}$, which
can be made arbitrarily large.
Results for the exponent $\tau$ from simulations of the BS branching
process are shown for one and two dimensions in Fig. \ref{six}.

The equivalence of the off-critical
BS branching process at a given value of $f_o$
with an $f_o$- avalanche in the stationary state of
the evolution model also proves that $f_o$-
avalanches in the stationary state of the evolution model have the same
statistical properties as the $G=f_o$- avalanches during the transient;
in particular they have the same probability
distributions.  The gap equation (\ref{e.approach}) maps the
transient, self-organizing behavior at a given value of the gap, $G$,
to the stationary $f_o=G$- avalanche distribution.  We emphasize,
again, that the $\gamma$ for the stationary distribution in the evolution
model, if it exists (i.e. if the scaling ansatz is valid),
 is the same $\gamma$ that enters into the solution of the gap
equation for the transient behavior of the  model.

For the other models, we propose analogously that the $G=f_o$-
avalanches during the transient also have the same scaling properties
as the $f_o$- avalanches in the stationary state.  The
picture is that the scaling behavior of
 $f_o$- avalanches is not affected by  correlations
at distances larger than the correlation length $\xi \sim (f_c -
f_o)^{-\nu}$ set by $f_o$.   When the self-organizing system
reaches a gap $G=f_o$, it has organized itself up to a scale $\xi(f_o)$; at
length scales smaller than this scale, it behaves as a critical
system.  Thus the exponent $\gamma$ that appears in Eqs. (\ref{e.gamma},
\ref{e.gamma2}) is the same $\gamma$ that enters into the transient
approach to the critical attractor, Eqs. (3-5).

\subsection{ The ``Gamma'' Equation}

We now proceed  to establish a general relation
for the  number of sites covered by an avalanche as the critical state is
approached. The relation is exact for the Sneppen and
evolution models, and a similar exact relation holds for IP.
It leads directly to a scaling relation between
the exponents $\tau$ and $\gamma$
 valid for all
extremal models. This  scaling relation
was previously derived for separate
models in \cite{ gamma}, \cite{ mppre},
\cite{ porous}, \cite{ narayan.93}, \cite{ olami.94}.
We will reproduce here in more detail the method of derivation used in
\cite{ gamma}, which gives not only the
scaling relation but the exact values of coefficients in it.
The following argument applies specifically to the evolution
 and Sneppen models;
we then generalize it to the other models.

Suppose that a value for the parameter
 $f_o<f_c$ has been chosen. The moments in time, $s$,
when the global minimal number $f_{min}(s)$ exceeds $f_o$
serve as  breaking points dividing the temporal axis into
a series of $f_o$ -avalanches, following one after another.
The probability that at any given moment the signal $f_{min}(s)$
will be greater than $f_o$ is given by
\begin{equation}
 p(f_{min}>f_o)=1/\langle S \rangle_{f_o}  ,\quad {\rm for} \qquad
f_o < f_c \qquad ,
\label{e.threshold}
\end{equation}
where $\langle S \rangle_{f_o}$ is the average size of an $f_o$- avalanche.

Let $n_{cov}$  denote the number of sites covered by an avalanche.
These sites had their
random number changed at least once during the course of the avalanche.
Since each site can be updated more than once, $n_{cov}$
is a priori different from the avalanche size $S$.  In fact, for
any avalanche
\begin{equation}
\label{ inequality}
n_{cov} \leq A S \qquad ,
\end{equation}
where $A$ is a constant which depends on dimension.

We can relate the divergences of these two physical quantities as the
critical state is approached.  This is accomplished by noting that, at
the moment in time when
 an $f_o$- avalanche is
finished, the random numbers at all  $n_{cov}$ sites
covered by the avalanche are
{\it uncorrelated} and {uniformly distributed in the
interval [$f_o$,$1$]}.  This is because at the {\it last update for each
site} within the avalanche the random number was chosen from an uncorrelated,
uniform distribution
between [0,1] {\it with the condition that it be larger than $f_o$.}
  We will make repeated use of this fact in what
follows.

Recall that when the $f_o$- avalanche started, the smallest random number
in the system was larger than $f_o$.  For simplicity, assume that
this number was also larger than $f_o + df_o$, where $df_o$ will
be taken to be vanishingly small.  When the $f_o$- avalanche finishes,
each of the $n_{cov}$ sites has equal
 probability $df_o \over 1-f_o$ to fall within the
interval $df_o$ above $f_o$. Therefore, $n_{cov}{df_o \over 1-f_o}$
is the probability that at least one of the $n_{cov}$ sites left behind by
the $f_o$- avalanche has a random number between
$f_o$ and $f_o+df_o$.
If now the auxiliary parameter $f_o$ is
raised by an infinitesimally small amount $df_o$, the breaking points
that had $f_{min}(s)$ between $f_o$ and $f_o+df_o$ will no longer stop
the $f_o+ df_o$- avalanches, and the average avalanche size will
increase.  This property of the avalanche hierarchy is demonstrated  in
Fig. \ref{seven}.
Consider a very long temporal sequence $f_{min}(s)$ which is also a sequence
of $N$, $f_o$ -avalanches.  Increasing the auxiliary parameter
to $f_o + df_o$, the number $N$ will be changed by
$dN=-N<n_{cov}>{df_o \over 1-f_o}$, to leading order in
$df_o$  for $N \rightarrow \infty$.
Of course, the sum of the  temporal durations of
these avalanches will remain constant.
This leads to the following differential equation for
the average size of an avalanche;
\begin{equation}
 {d\ln \langle S \rangle_{f_o}  \over df_o}
={\langle n_{cov} \rangle_{f_o} \over 1-f_o } \qquad .
\label{e.g1}
\end{equation}
This equation is exact for
the evolution and Sneppen models in any dimension.  It does not
require the use of any scaling assumptions.

In order to proceed further, we will now
 assume that the avalanche distribution
obeys the scaling ansatz, Eq. (\ref{e.av.dist}).  Then
for $f_o$ close to the critical value $f_c$,
the average size of the avalanche diverges as $(f_c-f_o)^{-\gamma}$
(Eq.\ (\ref{e.gamma})).
Substituting this power law into Eq. (\ref{e.g1})
gives
\begin{equation}
 \gamma=_{\lim f_o \rightarrow f_c} {\langle n_{cov} \rangle_{f_o}
(f_c-f_o) \over 1-f_o}
\qquad .
\label{e.g2}
\end{equation}
If the critical exponent $\gamma$ exists, then Eq. (\ref{e.g2}) is also
exact.  Note that the quantity $\gamma$ appears as a constant
rather than a critical exponent.  It
represents the average number of random numbers
between $f_o$ and $f_c$, left behind by an $f_o$- avalanche that has died,
in the limit that $f_o \rightarrow f_c$.
This number is universal.
For example, it does not depend on whether one updates
just nearest neighbors or also includes
next nearest neighbors.

A plot of $(1-f_o)/\langle n_{cov} \rangle_{f_o}$, for different values of
$f_o$, as shown
in Fig. \ref{eight} for the two dimensional evolution model, gives $f_c$ as
the intersection with the horizontal axis, and $1/\gamma$ as the
asymptotic slope close to $f_c$.
This enables us to determine $f_c$ very accurately, because
the uncertainty in the measurement of $\langle n_{cov} \rangle $ and
$\gamma$ leads to
an uncertainty not in $f_c$ but in $\Delta f$. Choosing a value of
$f_o$ near the critical point, $\Delta f=f_c - f_o$ is estimated from
Eq. (\ref{e.g2}). Specifically, in the one dimensional evolution model we
choose $f_o = 0.665$ and measure the average number of covered sites
$\langle n_{cov} \rangle_{0.665} =
446.9 \pm 7.0$.
Using this method for the evolution model,
we find
$f_c = 0.66702 \pm 0.00003$ for $d=1$ and $f_c = 0.328855 \pm 0.000004$
for $d=2$.  Our one dimensional result agrees with
Grassberger \cite { grassberger}.
As a by-product, the one dimensional result
rules  out the early speculation that $f_c$ is
exactly 2/3 in one dimension.
The exponent $\gamma$ can also be estimated accurately from points further
away from $f_c$.

Figs. \ref{nine} and
\ref{ten}  show the same plots for the one dimensional evolution model
and the one dimensional Sneppen model.   For the
one dimensional evolution model, our result,  $\gamma = 2.70 \pm 0.01$,
agrees with that
of Jovanovic {\it et al} who measured
$\gamma = 2.7 \pm 0.1$ \cite{ stanley} and Grassberger, who also measured
$2.71 \pm 0.03$ \cite{ grassberger}.

For the one dimensional Sneppen model
with $L=10^4$,
we find $f_c = 0.465$ and $\gamma \simeq 2.13$.  This value differs
substantially from that obtained by Sneppen $\gamma = 2.05 \pm 0.01$ at
$f_c = 0.46136 \pm 0.00005$
at a system size $L=5\times 10^5$ \cite{ gammasneppen}.
Our probably less accurate result is
due to finite size effects.
This value $f_c=0.465$ for the
Sneppen model also does not agree with the value of $f_c=0.4614$
that we determine from the avalanche distribution (see Section IV (D))
for a larger system size $L=2\times 10^5$.

For the evolution model,
the value of $f_c$ obtained using this new method, based on the
``gamma'' equation, is used as input  to determine
$\gamma$
from the divergence of the average avalanche size,
as $f_o$ approaches $f_c$, Eq. (\ref{e.gamma2}).
Our results for $\gamma$ for the evolution model using this
conventional method (this was the method used
in Refs. \cite{ stanley} and \cite{ grassberger})
are shown in Figs. \ref{eleven}, \ref{twelve}.
The two different methods we used
to measure $\gamma$ give consistent results for the evolution model.
Having measured $\gamma$, we can calculate the exponent $\rho$ for the
relaxation to the critical state (Eq. (\ref{e.trans})).  We find
$\rho={1/(\gamma -1)}=0.588 \pm 0.004$ and $\rho=1.43 \pm 0.01$ in the
one and two dimensional evolution models, respectively.

The scaling form, Eq. (\ref{e.av.dist}), for the divergence of
the average avalanche size
can be substituted into Eq. (\ref{e.threshold}), which becomes
$p(f_{min}>f_o) \sim (f_c-f_o)^{\gamma}$, or alternatively,
\begin{equation}
 p(f_{min}=f_o) \sim (f_c-f_o)^{\gamma-1}\qquad .
\label{e.threshold1}
\end{equation}

{}From Eq. (\ref{e.g2}) it follows that  the
average number of sites
$\langle n_{cov} \rangle_{f_o}$ covered
by an $f_o$ avalanche scales near the critical point as

\begin{equation}
\langle n_{cov} \rangle_{f_o} \sim (f_c-f_o)^{-1} \qquad .
\label{e.ncov}
\end{equation}
Since Eq. (\ref{ inequality}) holds for any avalanche, it is immediately
clear that $\gamma \geq 1$ for any model which obeys Eq. (\ref{e.ncov}).
This posteriorly confirms that Eqs. (\ref{e.trans}, \ref{e.trans2})
 are the only allowed possibilities
for the approach to the critical attractor via the gap equation.
The case $\gamma <1$ is not physically realized by the models considered
here.

We are now in a position to consider the spatial extent of avalanches.
Since the interactions are via nearest neighbor,
it is clear that all sites visited by an avalanche form a
connected cluster. In one dimension any connected cluster
is also compact.  Therefore, in any one-dimensional
model $n_{cov} \sim R_{cov}$, where $R_{cov}$ is the length of the
spatial interval
formed by covered sites.

In analogy with percolation,
it is conventional to define the characteristic spatial size
$R$ of an avalanche cluster as the mean square root deviation of the
set of all active sites in the avalanche from their center of mass.
In this definition each site is counted with the weight given
by the number of times it was visited by the avalanche. The avalanche
mass dimension $D$ is defined by the scaling relation
\begin{equation}
S \sim R^D \qquad,
\end{equation}
connecting the avalanche size $S$  (temporal duration) to its
spatial extent $R$.
In the case of a {\it compact} $d$-dimensional set of covered sites,
we can use another definition of the avalanche spatial size;
\begin{equation}
R_{cov}=n_{cov}^{1/d} \qquad .
\end{equation}
Assuming the absence of multifractal spatial scaling behavior for
avalanches implies that
\begin{equation}
R_{cov} \sim R \qquad .
\end{equation}

In all extremal models that we have studied numerically
thus far (excluding IP which is discussed at the end of this
section), the avalanche
mass dimension $D$ was measured to be
larger than the dimension of space $d$,
and the spatial projection of all active sites within the avalanche
was observed to form a compact object of dimensionality $d$.
In what follows, we assume
this is true for every extremal model below its upper
critical dimension.  Above the upper critical
dimension (provided that it exists) the fractal dimension of the collection
of covered sites is given by its mean field value $d_{uc}$,
and can no longer form a compact object in the space
of dimensionality $d>d_{uc}$.  Thus in all of the scaling relations
that include dimensionality,
derived below and summarized
in Table \ref{ scalingrelations}, $d$ should be replaced with
$d_{uc}$ for $d\geq d_{uc}$.  This behavior is somewhat analogous to
hyperscaling relations in equilibrium critical phenomena.

In the case compact avalanches
\cite{ commenta},
$n_{cov} = R_{cov}^d \sim R^d \sim S^{d/D}$ and from
Eq. (\ref{e.g2}) it follows that
$ (\Delta f)^{-1} \sim
 \langle n_{cov} \rangle
 \sim \int_{1}^{\Delta f^{-1/\sigma}} S^{d/D}S^{-\tau}dS
 \sim \Delta f^{(\tau-d/D-1)/\sigma}$.
As a result,
\begin{eqnarray}
 \sigma=1+d/D-\tau \label{e.sigmaBS} \qquad , \\
 \gamma={(2-\tau) \over (1+d/D-\tau)} \qquad .
\label{e.gammaBS}
\end{eqnarray}
The spatial correlation length exponent $\nu$
describes the scaling of the  cutoff in the spatial extent of the
avalanches, $r_{co} \sim \Delta f^{-\nu}$.  Since
$s_{co} \sim r_{co}^D$,
\begin{equation}
\nu=1/\sigma D=1/(D+d-D\tau) \qquad .
\label{e.nuBS}
\end{equation}

It is interesting to observe that Eq. (\ref{e.sigmaBS}-\ref{e.nuBS})
imply that the self-organization time to reach  the critical attractor
is independent of the initial state
of the system, for almost any initial condition.
A system
of size $L$ reaches the stationary state when $(\Delta f(s))^{-\nu} \sim L$.
The time $s_{org}$ required for
this scales as $s_{org}\sim L^{\tilde D}$, where
$\tilde D = d+{\gamma-1 \over \nu}$ (Eq. \ref{e.tildeD}).
Substitution of Eqs.\ (\ref{e.sigmaBS}-\ref{e.nuBS})
into this expression for $\tilde D$
gives the very simple result $s_{org}=L^D$, or $\tilde D = D$.
Instead of starting with the initial condition where the random numbers are
uniformly distributed in the interval [0,1], as
assumed in the gap equation (Eq. \ref{e.approach}), we may start
the organization process in a state where
all $f$'s are set equal to $1$ except one  site  which has
the value of $f$ equal to, say, $0.99$.  In this case,  none of
the original $1$'s are ever selected as the minimal site.
The organization process is finished at the moment
when the last $1$ is destroyed. This clearly  takes
$\sim L^D$ time steps, the same as for the transient that is
governed by the gap equation.

We checked these scaling relations numerically.
Figs. \ref{fourteen}a, \ref{fifteen}a show measurements of the
avalanche dimension $D$ for the one and two dimensional evolution model.
At the critical point $f_c$, $R_{cov}$
was measured as a function of $S$ for running avalanches
and their subavalanches in the
BS branching process, and averaged over more than $10^{11}$
updates.
In $d=1$, $D = 2.43 \pm 0.01$, and
in $d=2$, $D = 2.92 \pm 0.02$. 
These values are consistent with the scaling relation
Eq. (\ref{e.gammaBS})
and our measured values for $\tau$ and $\gamma$.  In fact,
Jovanovic {\it et al} \cite{ stanley} measured
the exponent for the divergence of the
average spatial size of the avalanche to be $0.98 \pm 0.03$
($\tilde{\nu}_{\perp}$ in their notation) for the one
dimensional evolution model. Taking into
account that in one dimension the spatial size
is simply proportional to $n_{cov}$ this
result is consistent with the exact value of $1$ we
predict.   In one
dimension, Eq. (\ref{e.nuBS}) can be written as
\begin{equation}
\tau -1 = \sigma \nu - \sigma
\label{e.taurelation}
\qquad .
\end{equation}
For the one dimensional evolution model, the results from Ref. \cite{ stanley},
$\tau = 1.08 \pm 0.05$, $\sigma=0.35 \pm 0.02$ and
$1/D= \sigma \nu = 0.43 \pm 0.01$ are consistent with
this last relation.  Similarly, Grassberger's results
$\tau = 1.073 \pm 0.003$, $\sigma = 0.343 \pm 0.004$ and
$1/D=\sigma \nu = 0.4114 \pm 0.002$ \cite{ grassberger}
are also consistent with
Eq. (\ref{e.taurelation}).

 This agreement occurs in spite of the
fact that it is very  difficult to determine the
true asymptotic value of $D= 1/(\sigma \nu)$.
One reason is that the apparent exponent may be nonmonotonic, as demonstrated
in Fig. \ref{fourteen}b.  The effective exponent as plotted
here is measured by numerically calculating the local slope of
the curve $R$ vs. $S$.  It is apparent for the two dimensional model that
the effective exponent, $\sigma \nu=1/D$,
 does not approach its asymptotic value monotonically
but first overshoots  with a maximum around $S \simeq 10$ and
then undershoots  with a minimum around $S \simeq 10^2$.
There are indications
that similar behavior may occur
in the one dimensional model at much larger avalanche scales.
Our numerical capability for the one dimensional model is not
sufficient to be sure that we are observing   asymptotic behavior
despite studying $3\times 10^{11}$ time steps, and thus $3\times
10^{11}$ avalanches and sub-avalanches.
In addition, simulations of the related $M \rightarrow \infty$ evolution
model \cite{ boettcher}, where $D=4$ is known exactly, indicate that
there are logarithmic corrections for $R_{cov}$.
Thus,  the numerical
studies thus far have not conclusively
refuted the conjecture \cite{ gap}, \cite{ gamma},
based on symmetry arguments, that the
exponents $D$ and $\tau$ in the evolution are the same as
directed percolation, or Reggeon Field Theory.  It seems dangerous
to us to base conclusions about universality on differences of
less than two percent in the observed ``asymptotic'' regime
at $S\simeq 10^6$ \cite{ grassberger} in
one dimension.
In two dimensions, the disagreement is within error bars and less than
one percent.

Eq. (\ref{e.g1})
can be generalized to include
IP with the following consideration.
In IP the list of growth sites
from which the minimal number is selected
is not a compact $d$-dimensional
lattice of linear
size $L$, but the fluctuating fractal boundary of the invaded
cluster.
Comparing the list of potential growth sites before
and after an $f_o$- avalanche,
one observes that some  sites
were invaded during this specific
avalanche and thereby removed from the  list.
In addition, the result of an $f_o$-avalanche in
IP-2 and IP-3 may be
a new internal ``lake'' in the invading cluster, so
sites on the lake perimeter are now excluded from the active boundary as well
(see Appendix).
On the other
hand, some new boundary sites may have been added.  We will denote this
latter number
of {\it new} boundary sites added by an avalanche of size $R$ as
$n_{new}(R)$.  Since these sites constitute a part of
the active boundary of the region invaded by the avalanche,
it is reasonable to assume that
\begin{equation}
\label{e_n_new}
  n_{new}(R) \sim R^{d_B} \qquad ,
\end{equation}
where $d_B$ is the fractal dimension of active boundary of the invaded
region.
All of these new random numbers
are uncorrelated and uniformly distributed between
$f_o$ and $1$.  By replacing the number of
covered sites, $n_{cov}$, with the number of new sites added to
the boundary, $n_{new}$, we find that
Eqs. (\ref{e.g1}-\ref{e.g2})  apply to IP as well.

Using Eq. (\ref{e_n_new}) we can
derive exponent relations analogous to Eqs. (\ref{e.sigmaBS}-\ref{e.gammaBS})
for IP:
$(\Delta f)^{-1} \sim
 \langle n_{new} \rangle_{f_o}
\sim \int _1^{\Delta f^{-\nu D}}
 s^{d_B/D}s^{-\tau }ds
\sim \Delta f^{(\tau -1-d_B/D)\nu D}$, which gives
\begin{equation}
 \tau=1+{d_B-1/\nu \over D} \label{e.tauIP}\qquad ,
\end{equation}
\begin{equation}
 \gamma=1+(D-d_B)\nu  \label{e.gammaIP} \qquad .
\label{e_gexp_ip}
\end{equation}
Although the exact equations (\ref{e.g1}-\ref{e.g2}), with $n_{cov}$
replaced with $n_{new}$,
are new for invasion percolation,
the scaling relations (\ref{e.tauIP}-\ref{e.gammaIP}) are well known.
To the best of our knowledge, they were first derived by Gouyet
\cite{ gouyet}.
Substitution of  Eqs. (\ref{e.gammaIP}) and
(\ref{e.boundary}) into  Eq. (10)
simplifies the final form of the transient approach to the critical state
for IP:
\begin{eqnarray}
\label{e.ipdeltaf}
\Delta f \sim (s/L^d)^{-{1-g \over \gamma-1}}\nonumber \\ \sim
(s/L^d)^{-{1-g \over \nu(D-d_B)}} \sim (s/L^d)^{-{1 \over \nu(D-d)}} \quad .
\end{eqnarray}
The time required to complete self-organization $s_{org}\sim
L^{d+{\gamma-1 \over \nu(1-g)}}\sim L^D$.
Eq. (\ref{e.ipdeltaf})
 agrees with results from \cite{ rouxguyon} obtained using
different arguments.

Eq. (\ref{e.threshold}) holds for any extremal
model where $f_{min}(s)$ has a well defined threshold level $f_c$ in the
thermodynamic limit. At the same time,  though,
Eqs. (\ref{e.g1}-\ref{e.g2}) are not valid for the Zaitsev
and LIM models.
The reason is that
for these two models the distribution
of internal forces left by an $f_o$- avalanche is not flat and uniform
in the interval
$[f_o,1]$.  However, as long as this distribution is not singular in the
limit that $f_o \rightarrow f_c$, and these internal forces are
uncorrelated, as is reasonable to assume,
the result $\langle n_{cov} \rangle \sim (f_c-f_o)^{-1}$
still holds, and, therefore, the scaling relations
(\ref{e.sigmaBS}-\ref{e.nuBS}) are valid.
For  interface models, it is customary to express scaling
relations
in terms of $D$, $\nu$ as basic exponents.
Our scaling relations will then read
\begin{eqnarray}
\tau=1+{d-1/\nu \over D} \qquad ;
\label{e.int.tau} \\
\gamma=1+(D-d)\nu  \qquad .
\label{e.int.gamma}
\end{eqnarray}

\subsection{ Law for Stationary States: $\eta =0$}

In this section we derive a law for the activity in the stationary state for
the class of SOC extremal models defined in Section II.
 Recall that an $f_c$- avalanche starts when {\it all} sites in the
system have random numbers, or internal forces, above the threshold
$f_c$.  In the following discussion, we will only consider $f_c$- avalanches,
and therefore drop the label $f_c$.  During the course of the avalanche
there will be sites, denoted as {\it active} sites, or ``particles'',
where the random numbers $f_i$ are less than the threshold
$f_c$.  The number of these active sites after $s$ updates is denoted
 the ``activity'', $n(s)$.  This quantity is just the number of sites
in the system that have random numbers less than $f_c$.
The current avalanche stops when all sites in the
system again have random numbers above $f_c$,
or when $n(s)$ first returns to zero.  Then a new avalanche will start
somewhere else in the system.
The  probability distribution for
avalanche sizes, $P(S) \sim
S^{-\tau}$, is therefore, by definition, also
the probability distribution of
interval lengths between subsequent returns of $n(s)$ to zero.

Fig. \ref{sixteen} shows a typical $n(s)$.  The collection of return points,
$\{n(s)=0\},$ forms a fractal with dimension ${\tilde d}= \tau -1$
on the one-dimensional time line.  This relationship
for $\tilde d$ is explained in
the next section (see Eq. (\ref{e.tau_f})).
As long as $\tilde d < 1$,  the threshold $f_c$ is well defined in
the thermodynamic $L \rightarrow \infty$ limit.
The initiation of a new avalanche in the stationary
state can be viewed as the injection
of a single ``particle'' into the system.
Thus, the
 average number of return points, or number of injected particles, added in
an interval of $s$ steps is $n_{INJ} \sim s^{\tilde d}$.

We examine the average number of active sites,  $S$ steps after
a {\it single} particle has been injected into the system, $<n(S)> \sim
S^{\eta}$.   This quantity is by definition
the average activity $n_{SURV}(S) \sim S^{d_s}$
 of the avalanches that survive $S$ steps multiplied
by the probability of survival, $P(s'>S) \sim S^{1-\tau}$.
Avalanches that die before $S$ steps are counted with
{\it zero} particles in the average $<n(S)>$, while the surviving
avalanches are counted with their actual number of active sites
$n(S)$.  The quantity
$n_{SURV}(S)$ is an average over  avalanches that
survive $S$ steps, while $<n(S)>$ is an average over {\it all} avalanches,
hence the factor $P(s'>S)$.
We now utilize the hierarchical nature
 of the  stationary process: we propose that
the activity of the large avalanches after $S$ steps scales in the
same way as the amount of activity injected into the system during
a time interval of length $S$:

\begin{equation}
n_{SURV}(S) \sim n_{INJ}(S) \ , \ \  \ {\rm or} \ d_s = {\tilde d}= \tau -1 \ .
\label{e.nact}
\end{equation}
That is, there is only one dimension for the activity in the critical
process.
Thus,
\begin{eqnarray}<n(S)>\; \equiv n_{SURV}(S)\; P(s'>S)\nonumber \\
 \sim n_{INJ}(S) S^{1-\tau} \sim S^0
\ ; \ {\rm i.e. } \ \ \eta=0 \qquad .
\label{e.eta}
\end{eqnarray}

In order to illustrate the argument by an elementary example, consider
a simple, uncorrelated one dimensional
 random walk, with the activity increasing or decreasing by
one with equal probability
at each step.  This is the mean field, SOC state in many
models
\cite { flyvbjerg}, \cite{ rayjan}, \cite{ stonybrook}.
The infinite random walk can be viewed as a
sequence of avalanches.  These avalanches are, by definition,
 the intervals between
subsequent returns of the walk  to zero.
The exponent $\tau$ for the first return times is $\tau=3/2$,
so the probability of
$s'$ exceeding $ S$ scales as $S^{-1/2}$.
The average number of  returns in $S$ steps, $n_{INJ}(S)\sim
S^{1/2}$, scales in the same way as
the activity of surviving avalanches after $S$ steps, $n_{SURV}(S)
\sim S^{1/2}$.
Thus, the
average activity after $S$ steps of a single
avalanche  is $<n(S)> \sim S^{1/2}S^{-1/2} \sim S^0$, and $\eta = 0$
rigorously.

Fig. \ref{seventeen} shows the average activity of a single
avalanche, $<n(S)>$, in the
two dimensional  evolution model.
The exponent $\eta$  reflects the average activity
$S$ time steps after starting a single
realization of the BS branching process at $f_c$.
The quantity $<n(S)>$ saturates for large $S$,
varying only 2\% over {\it four decades};
this is consistent with $\eta = 0 \pm 0.002$ in  two
dimensions.  Fig. \ref{seventeen}
 is our strongest numerical confirmation of the
$\eta = 0$ prediction.  As eluded to earlier, c.f. Fig. \ref{fifteen}b,
in one dimension there are larger corrections to scaling
and the saturation regime obtained numerically is smaller,
although the overall behavior is the same, as shown in Fig.
\ref{eighteen}.  The slope is decreasing and reaches $\eta = 0.01 \pm 0.02$
at $s=6 \times 10^7$.

The fractal dimension of active sites within a surviving
avalanche is defined by the relation
$n_{SURV}(r) \sim r^{d_f}$, where $r$ is the spatial extension of
the avalanche cluster.
Using the relation $s \sim r^D$,  and Eq. (\ref{e.nact}),
we find the  relation
\begin{equation}
d_f = D d_s = D(\tau -1) \ ,
\label{e.fractald}
\end{equation}
connecting the geometrical properties of the active sites to the
properties of the entire
avalanche clusters. Fig. \ref{seventeen} (\ref{eighteen})
  also shows
$n_{SURV}(S)$ vs. $S$ in two (one) dimensions for the evolution model.
At the largest time scales measured,  the slope of the curves yields
$d_s = 0.11 \pm 0.02$ and
$d_s = 0.25 \pm 0.005$
in one and two dimensions. In two dimension, this
is consistent with our measured value of $\tau-1$ (c.f. Fig. \ref{seven}),
but in one dimension our measured value of $\tau-1$ is somewhat lower. This is
probably a consequence of the slow convergence of the slope, as is evident
from Fig. \ref{eighteen}.  Actually, Grassberger measured $n_{SURV}(S) \sim
S^{0.74}$ for the one dimensional evolution model, which agrees
with his measured value for $\tau$
\cite{ grassberger}, and thus with Eq. (\ref{e.nact}) and $\eta=0$.

The usual dynamical exponent relating space and time,
$z$, where $t \sim r^z \sim s^{z/D}$, can also be measured.
The parallel time unit, $t$, is different than the sequential
time unit $s$.  It is defined as
the average number of update steps for changing all active sites,
so that time increases by an amount $1/n(s)$ at each update;
i.e. $s \rightarrow s+1$, $t \rightarrow t + 1/n(t)$.
It appears naturally in models for depinning at constant force,
for example, because all unstable sites move together.  We introduce
the dynamical exponent $z$ also for the sequential models in order
to compare dynamical critical behavior in the different situations.

For this definition of time, clearly $s \sim n t$
and $d_s+z/D=1$.  Within the superuniversal class where
$\eta=0$, including the extremal SOC models discussed here,
the exponent $z$ is given by the scaling relation
\begin{equation}z=D - d_f= D(2-\tau) \ .
\label{e.z}
\end{equation}
We
measured the dynamical exponent $z = 1.19 \pm 0.05$
for the Sneppen model in one
dimension as shown in Fig. \ref{twenty}.  This value is close
to the  theoretically predicted value  $z=1.21$, obtained by using
Eq. (\ref{e.z}) with $D=1.63 \cite{ mppre}$, and our measured value of $\tau$.
We have also measured the dynamical exponent $z$ for the one
and two dimensional evolution models as shown in Fig. \ref{twentyone}.
In two dimensions, $z = 2.16 \pm 0.02$.
 Inserting our measured values for
$\tau$ and $D$ into the expression for $z$  we predict $z = 2.20 \pm
0.04$, in fairly good agreement. In one dimension we find $z = 2.26 \pm 0.05$
to be compared with the somewhat lower measured value $ 2.10
\pm 0.05$.  This discrepancy again may reflect slow
convergence for the one dimensional evolution model.
Jovanovic {\it et al} \cite{ stanley}
measured
a dynamical exponent (in their notation $\nu_{\parallel}/\nu_{\perp}$)
$=2.1 \pm 0.1$
for the one dimensional evolution model using the tree structure of the
avalanches.  The dynamical exponent that they measured may be equivalent
to our definition of $z$, and their result is consistent with our
prediction, again within numerical uncertainty.

An alternative way of deriving $\eta =0$
follows:
Consider either the evolution or Sneppen model in the stationary state,
 and select an arbitrary
value  $f_o$  below
$f_c$.
We consider the number of active sites with random numbers
below $f_c$.
  As explained in the preceding subsection,
when an $f_o$- avalanche ends, it leaves behind
uncorrelated random numbers uniformly distributed between
$f_o$ and one.  As a result, the average number of
active sites created, or left behind,  by an $f_o$- avalanche is
exactly \  $ <n_{cov}>_{f_o } (f_c-f_o)/(1-f_o)$ (see Eq. (17)).
We assume that when the $f_o$- avalanche started, the
active sites between $f_o$ and $f_c$ were
distributed on a fractal,
with dimension $d_f$.  As a result, the average
number of actives sites  that are
destroyed by the $f_o$- avalanche
scales as $<R^{d_f}>_{f_o}$, where $R$ is the spatial
extension of the avalanche.
In the stationary, critical state
the average number of active sites
created and destroyed by an
$f_o$- avalanche must be equal.  Assuming a power law distribution
of avalanches sizes with cutoff,
$R_{co} \sim (f_c-f_o)^{-\nu}$,
 we find
\begin{equation}
1=\nu (d-d_f)       \ .
\label{e.hyper}
\end{equation}
Using the scaling relation,  $\tau =1+(d-1/\nu)/D$ (Eq.
(\ref{e.nuBS}))
we again obtain our fundamental result $\eta=d_f/D-\tau+1 =0$.
Eq. (\ref{e.hyper}) is reminiscent of the hyperscaling relation
$\beta = \nu(d-d_f)$ valid for equilibrium  systems, and
for directed percolation \cite{ Grassb79}.

In order to take this alternative argument over to the Zaitsev model and
LIM, one makes the same assumption that was made for
the gamma equation, i.e.  the distribution
of internal forces left behind by an $f_o$- avalanche is uncorrelated
and smooth.
We have measured $<n(s)>$ and $n_{SURV}(s)$ for the Zaitsev flux
creep model
\cite{ zaitsev} in one dimension, as shown in Fig. \ref{twentytwo}.
These results are consistent with $\eta =0$.
Finally, we predict that $\eta =0$ for invasion percolation.

 Although the $\eta=0$ law applies to a broad class of SOC phenomena,
it does
not generally apply to systems that are tuned to be critical, since
for non-SOC systems the critical
state may not be
 stationary.  For instance,  in directed percolation $\eta= d_f/D
- \tau + 1 \simeq .21$ \cite{ remark}
 in one dimension.   Also,
in directed percolation
the fractal dimension of active sites obeys a hyperscaling
relation, $d_f = d - D(\tau - 1)$.
This hyperscaling relation is inconsistent with the relation
$d_f = D(\tau -1)$ that holds for the extremal SOC models that we study.
Thus directed percolation does not belong to the superuniversal
class defined by the dimension independent law $\eta=0$.

The models for depinning
at constant force, such as TLB or parallel
LIM, also do not in general obey $\eta =0$, or the corollary results
$d_f=D(\tau -1)$, $z=D(2-\tau)$.  For example, in the one
dimensional TLB at constant force,
 it has been predicted and confirmed numerically
that $z=1$ \cite{ paralleltl}.  Our measured value for
this model $z=1.00 \pm 0.05$, as  shown in Fig. \ref{twenty},
is consistent with this prediction.
Since $D \simeq 1.63 $ and $\tau \simeq 1.26$
are the same as for the Sneppen model (see section VI), one
can compare with the different prediction $z \simeq 1.21$
based on the $\eta=0$ law.
Since $\eta$ is not zero for the tuned depinning models
at constant force, while
it is zero for the SOC depinning models, the critical dynamics of a
self-organizing system are different than the
critical dynamics of a tuned system.
  Thus SOC cannot be simply viewed as sweeping an instability
\cite{ sornette}.  If the TLB model were to be studied at constant
{\it velocity}, we predict that $\eta =0$ and $z \simeq 1.21$, as for
the SOC (Sneppen) case.  Thus the dynamical critical exponents,
such as $z$, depend on how the depinning transition is probed.

The existence of a stationary limit
may imply that $\eta=0$ for a very broad class of
SOC models, beyond those studied here.  In addition, Pietronero
\cite{ fst}
and co-workers explicitly make use of the stationary limit in the
Fixed Scale Transformation method.  This suggests that stationarity may
potentially provide more
powerful tools to understand SOC and fractal
growth phenomena in a wider range of systems.

\subsection{Backward Avalanches.}

The dynamics of extremal models forms a  hierarchical
structure.
We have defined $f_o$- avalanches as the activity between subsequent
moments in time when the signal $f_{min}(s)$
is larger than an auxiliary parameter $f_o$.
Then, as the parameter $f_o$ is decreased,
 larger avalanches are subdivided into smaller
ones because new breaking points appear.
For the sake of clarity in this section, we will now refer to
avalanches defined by this specific
rule as $f_o$-punctuating avalanches, since they
correspond to a sequence of events between subsequent punctuations
of the $f_o$-barrier by the signal $f_{min}(s)$.
Another way of looking at the hierarchy is to
define avalanches by a slightly different rule.
According to the new rule,
a  forward avalanche begins at {\it every time step $s$.}
It runs as long as $f_{min}(s+s') \leq f_{min}(s)$,
and will stop at the first moment $s+S$ {\it forward} in time
when $f_{min}(s+S)>f_{min}(s)$. This avalanche
is similar to an
$f_o$-punctuating avalanche, except that now we
set $f_o$ exactly to the value of the signal $f_{min}(s)$
at the starting point of the avalanche.
In the evolution model, the mapping to the BS branching process
proves that the
probability to have an $f_o$-punctuating avalanche of size $S$,
is {\it exactly} the same as the conditional probability to
have a forward avalanche of this size,
given that the value of the signal at the beginning of the
avalanche was $f_o$.
That is, the exact knowledge of the value of the signal
$f_{min}(s) \geq f_o$ at the starting point of an $f_o$-avalanche doesn't
influence its statistics. This follows from the observation
that the site carrying $f_{min}(s)$ gets
a new random number at the very first time step of the avalanche
and all the information about its previous value is erased from the system.
Therefore, the conditional probability $P_f(S,f_o)$  to have
a forward avalanche of size $S$, given that the signal
at its starting point was equal to $f_o$, is exactly equal to the
probability $P(S,f_o)$ to have an $f_o$-punctuating avalanche of size $S$.
Then from Eq. (\ref{e.av.dist}),
\begin{equation}
\label{e.forw}
 P_f(S,f_o)=P(S,f_o)= S^{-\tau }g(S(f_c-f_o)^{1/\sigma})
\qquad .
\end{equation}
For the other models we lack a rigorous proof,
 so we simply conjecture that both $P_f(S,f_o)$ and
$P(S,f_o)$ scale (e.g. Eq. \ref{e.av.dist}) with the same
exponents $\tau$ and $\sigma$, but with possibly different
scaling functions $g_f(x)$ and $g(x)$.  With this assumption, the
results which follow in this  section also apply to the other
SOC extremal models.

To study the properties of the signal under
time reversal it is useful to define {\it backward avalanches}.
Now we look for the first moment {\it back}
in time when
\mbox{$f_{min}(s-S)>f_{min}(s)=f_o$.}
The definitions of both forward and backward avalanches
are illustrated in Fig. \ref{fback}.
This figure  demonstrates
the hierarchy in the avalanche structure:
all forward and backward avalanches that start inside
one big forward
(backward) avalanche are constrained to not  go beyond the limits
of the parental avalanche and, therefore, can be considered
to be its sub-avalanches.
Each  sub-avalanche in  turn has its own sub-avalanches, and so on.
One can look at the entire activity in an extremal model as one great
parental critical avalanche, which began in the distant
past. It contains
sub-avalanches of all sizes.

Since all extremal processes are intrinsically irreversible,
it is possible to have
different statistical properties for the
forward and backward
avalanches.  In analogy with
Eq. (\ref{e.forw}), we conjecture
a scaling form for the conditional
backward avalanche probability distribution;
\begin{equation}
\label{e.back}
 P_b(S,f_o)={1\over N} S^{-\tau _b}g_b(S(f_c-f_o)^{1/\sigma})
\qquad ,
\end{equation}
 where $g_b(x)$ is a scaling function that rapidly decays to zero for
$x \gg 1$. We will prove later that the cut-off exponent $\sigma$
in this expression is the same for both forward and backward
avalanches, while the power law exponents
$\tau_b$ and $\tau$ are different. In fact
$\tau_b<1$, so that a
 normalization factor $1/N$ must be included in
the conditional probability distribution for backward avalanches.

According to Eq. (\ref{e.threshold1}),
the minimal numbers selected at different time steps are distributed
with probability density
\mbox{$p(f_{min}=f_o) \sim (f_c-f_o)^{\gamma-1}$,}
where $\gamma={2-\tau \over \sigma}$ is the critical exponent
governing the divergence of the average punctuating avalanche size.
The distribution of {\it all} forward avalanches,
$P_f^{all}(S)$ is obtained by integrating the
conditional probability from Eq. (\ref{e.forw}) with the proper weight
from Eq. (\ref{e.threshold1}) to give
\begin{eqnarray}
\label{e_forw_all}
 P_f^{all}(S)=\int_{0}^{f_c} P_f(S,f_o)\; p(f_{min}=f_o)df_o
\nonumber \\
=\int_{0}^{f_c} S^{-\tau}g(S(f_c-f_o)^{1/\sigma})
\; (f_c-f_o)^{\gamma-1}\; df_o
\nonumber \\
=S^{-\tau - \sigma \gamma}=S^{-\tau-\sigma(2-\tau)/\sigma}=S^{-2}
\quad .
\end{eqnarray}
The unexpected result is that
the exponent for this distribution has the {\it superuniversal}
value -2  in all dimensions.

For the distribution of all backward avalanches $\tau_b<1$, and
the normalization
factor $1/N = 1/\int_1^{\infty}S^{-\tau_b}g_b(S(f_c-f_o)^{1/\sigma})dS=
(f_c-f_o)^{1-\tau_b \over \sigma}$ enters.
Integrating Eq. (\ref{e.back})
gives
\begin{eqnarray}
\label{e_all_b}
P_b^{all}(S)=\int P_b(S,f_o)\; p(f_{min}=f_o)\; df_o \nonumber \\
\sim S^{-\tau_b - \sigma(1-\tau_b)/\sigma -\sigma(2-\tau)/\sigma}=
S^{\tau-3}
\qquad .
\end{eqnarray}
The exponent $\tau_b$ does not enter into the final
expression for $P_b^{all}(S)$ as long as $\tau_b<1$. The
exponent for the  distribution
of all backward avalanches is model and dimension dependent
and is
related to the usual $f_o$- avalanche distribution
exponent $\tau$.

In numerical simulations,
the exponent $\tau$ is conventionally measured from
the distribution $P(S,f_o)$ of $f_o$-punctuating avalanches
with the value of $f_o$ carefully chosen as close
as possible to the actual threshold $f_c$.
Since
backward avalanches are defined at every
time step, while $P(S, f_o)$ gets contributions only
when $f_{min}(s)>f_o$,
the distribution $P_b^{all}(S)$ has much better statistics
than $P(S,f_o)$ after the same number of time steps.
It is also very convenient  that $P_b^{all}(S)$  automatically has
no cut-off, so one
does not need to know  $f_c$ in order to measure
its power law.
Thus,  $P_b^{all}(S)$
provides, in principle, a very accurate way to
measure $\tau$ for the extremal SOC models discussed here.
 However, we have not yet pushed this technique to
its ultimate limit.

We have measured $\tau_b^{all}=3-\tau$ in the one dimensional Sneppen model,
and the one and two dimensional LIM. These results are shown in
Figs. \ref{f_ba_snep1} and \ref{f_ba_LIM1}.
For the Sneppen model, our numerical results from
the backward avalanche distribution
give  $\tau = 1.255 \pm 0.02$,
 in agreement with the value of $1.25 \pm 0.05$ measured
by Tang and Leshhorn \cite{ sneppentl}, and $\tau=1.26 \pm 0.01$
measured by us (Fig. \ref{sneppenavalanche})
using the conventional method.
The LIM results, $\tau_{1D}=  1.13 \pm 0.03 $
and $\tau_{2D}= 1.29 \pm 0.03$,
are in agreement with the results from numerical simulations
on self-affine one dimensional interfaces in porous media
from \cite{ martys}, and two dimensional domain walls in  the
Random Field Ising Model
\cite{ rfim}.   This supports the
possibility that these models might be in the universality
class of the LIM; a more detailed comparison is made in Section VI.

A study of backward avalanches
leads to an exponent relation for $\tau_b$, and a proof
that $f_o$-backward avalanches have the same cutoff as $f_o$-punctuating
(or forward) avalanches. For the sake of simplicity we concentrate
on the case of the evolution model, where the only assumption is that
 the scaling forms, Eqs. (38, \ref{e.back})
for the avalanche distributions exist.  With the additional assumptions
that have already been mentioned, the resulting scaling relations
also  apply to the other SOC extremal models defined in Section II.

Consider an arbitrary $f_o$-punctuating avalanche.
The probability distribution $P(S,f_o)$ for the
size $S$ of this avalanche is given by Eq. (\ref{e.forw}).
For this $f_o$- punctuating
avalanche to be also a valid $f_o$-backward avalanche,
starting at $s+S$ and running backwards in time down to $s$,
one needs $f_{min}(s+S)=f_o$.  We next calculate
what fraction of $f_o$-punctuating avalanches are also
$f_o$-backward avalanches.
Suppose we have a temporal sequence $f_{min}(s)$ which is
an ensemble  of $N$,
$f_o$-punctuating avalanches, where $N$ is a sufficiently large number.
The average number of $f_o$- punctuating
avalanches of size $S$ in such an ensemble
is given by $N(S)=NP(S,f_o)$.
At the end of any
$f_o$-avalanche of size $S$,
$n_{cov}$ sites have acquired new
random numbers.  If the avalanches are compact, then $n_{cov} \sim S^{d/D}$.
All these random numbers are {\it uncorrelated} and uniformly
distributed between $f_o$ and $1$.
To have $f_{min}(s+S)=f_o$ we need the minimal number in the system to lie
between $f_o$ and $f_o+df_o$. This number can be only at one of these
$n_{cov}$ updated sites, since
at the beginning of the avalanche every number in the
system was larger than $f_o$.
The probability that at least
one of these numbers will be between $f_o$ and $f_o+df_o$ is given
by $n_{cov}{df_o \over 1-f_o} \sim S^{d/D} {df_o \over 1-f_o} $.
The number $N_b(S)$ of valid $f_o$-backward
avalanches of size $S$ in our ensemble is
$N_b(S)df_o =NP(S,f_o) n_{cov}{df_o \over 1-f_o} \sim
NP(S,f_o) S^{d/D}{df_o \over 1-f_o}$.
Therefore, the conditional
probability distribution of $f_o$-backward avalanches
obeys:
\begin{equation}
\label{e_back}
P_b(S,f_o) \sim S^{d/D}P(S,f_o) \qquad ,
\end{equation}
where the proportionality constant is determined from the normalization
condition $\sum_{S=1}^{\infty} P_b(S,f_o)=1$.

Intuitively, it is clear that larger
 $f_o$-punctuating avalanches affect a larger number of sites.
  These larger avalanches
are more likely to leave behind a new random number between
$f_o$ and $f_o+df_o$ and thus constitute  a valid
$f_o$-backward avalanche. This explains why the
distribution of backward avalanches
acquires an additional factor of $S^{d/D}$ compared to
punctuating (or forward) avalanches.
Eq. (\ref{e_back}) shows that both forward and backward avalanche
distributions have the same cut-off as a function of $f_o$,
and their power law exponents obey the relation $\tau_b=\tau-d/D$.

{}From  Eq. (\ref{e.int.tau}), we get a particularly
simple expression for  $\tau_b$ :
\begin{equation}
\label{e_tau_b}
 \tau_b=1+{d-1/\nu \over D}-{d\over D}=
 1-{1 \over \nu D }=1-\sigma \qquad.
\end{equation}
Since $\sigma$ must be positive, this proves that
$\tau_b < 1$ and posteriorly confirms the validity of
Eq. (\ref{e_all_b}).

All equations in this section apply to
IP with the dimension $d$ replaced with $d_B$ (see Appendix A).
The importance of forward avalanches in IP
was first recognized by Roux and Gouyon \cite{ rouxguyon}.
However, they made some erroneous assumptions
which lead  to incorrect scaling relations.
Our exponent relations agree well with their numerical
results, though.
For IP-2, where $d_B=1.75$ and $D=1.89$,
they measured $\tau_b^{all}=1.50 \pm 0.04$ which is
consistent with our prediction $\tau_b^{all}=1.47$ based
on Eq. (\ref{e_tau_b}) and the assumption that $\nu=4/3$ as in
ordinary percolation \cite{ grimmett}.

\subsection{Levy Flight Distribution}

In the stationary state, the minimal site jumps throughout the system in
a correlated and anomalous fashion which has some similarity to the
usual Levy flight picture.
Specifically,
one can record the spatial location $\vec{r}_{min}(s)$ of the current
extremal site (with the smallest random number)
as a function of time $s$ \cite{ zaitsev},
\cite{ bsmodel}, \cite{ sneppen}.  The
distribution $P_{jump}(r)$ of jumps
$r=|\vec{r}_{min}(s)-\vec{r}_{min}(s-1)|$ between subsequent
extremal sites follows a power law:
\begin{equation}
\label{e_pi_def}
P_{jump}(r) \sim r^{-\pi} \qquad .
\end{equation}
This behavior is
reminiscent of the Levy Flight Random Walk (LFRW), where at
every time step the walker jumps
in a random direction by a distance that is drawn from a
power law distribution.  In contrast to the uncorrelated
LFRW, the jumps of activity in extremal models are temporally
correlated, so that even if $\pi >3$ and, therefore,
the jump distribution has
finite first and second moments, the process does not
necessarily reduce to
 ordinary diffusion.

The exponent $\pi$ can be related to other exponents
as follows:
Consider a backward avalanche that started at time step $s$. Suppose
it has size $S$.
By definition,
$f_{min}(s-S) > f_{min}(s)$, while $f_{min}(s-k) < f_{min}(s)$ for
$1 \leq k \leq S-1$.
Looking at the same sequence of events
forward in time from time step $s-S$ to $s$,
one notices that the forward avalanche
with $f_o=f_{min}(s-S)$ that was started at time step $s-S$
is still running at $s$. At this moment $s$,
$n_{cov} \sim S^{d/D}$ sites acquired  new random numbers
since time step $s-S$. The active sites at time steps
$s$ and $s-1$ are both rigorously constrained to belong to this
specific set
of covered sites.

Looking further back in time one may find another
forward avalanche, which is still running at time step $s$.
Say, it
started at time step $s-S'<s-S$  with
$f_{min}(s-S') > f_{min}(s-S)$, and has covered a larger spatial region
$R' \sim S'^{1/D}$.
Such an avalanche will contain the avalanche that started at time step
$s-S$ as one of its sub-avalanches.
Active sites at time steps
$s$ and $s-1$ will obviously belong to the bigger
spatial region of size $R'$ as well.
The importance of the backward avalanche
is that it automatically selects the {\it smallest} forward avalanche
containing both sites
$\vec{r}_{min}(s-1)$ and  $\vec{r}_{min}(s)$ and, therefore, imposes
the most restrictive constraint on their relative positions.

For the evolution model,
we  proceed by showing that: 1) the position of activity
$\vec{r}_{min}(s)$ at time step $s$ is
equally likely to be at one of the
$n_{cov}$ sites with new random numbers;
2) It is uncorrelated with the previous position of
the active site $\vec{r}_{min}(s-1)$, within the set of $n_{cov}$ sites.
To do this we will again use the powerful
observation that any $f_o$-avalanche of size $S$
leaves in its wake a set of uncorrelated random numbers.
For our avalanche in question we can set $f_o$  equal
to $f_{min}(s)$. Then $n_{cov}$ new numbers are uniformly distributed between
$f_o$ and 1 and are equally likely to host the current global minimum.
Looking at  the $n_{cov}$ sites at time step $s-1$,
just before the update was performed, one observes
that almost all of these sites have already acquired
their final uncorrelated number between $f_o$ and 1.
The only sites which can potentially be active and, therefore,
correlated are $\vec{r}_{min}(s-1)$
itself and its nearest neighbors. Because at the next
time step the $f_o$-
avalanche dies out,  all $2d+1$ new random numbers
created at this last time step must be larger than $f_o$.
They simply join the rest of the $n_{cov}$ sites, which already have their
random numbers between $f_o$ and $1$, and become indistinguishable
from them. Therefore, the position of the active site at time step $s$
is uncorrelated from the particular position, within the set
of $n_{cov}$ sites, of the active
site at time step $s-1$.
This finishes the proof  that the current jump of activity
$r=|\vec{r}_{min}(s)-\vec{r}_{min}(s-1)|$ for the evolution model
 is bounded only
by the linear size $R$ of the backward avalanche that  starts
at time step $s$.  We propose that this bound also holds for the
other extremal models considered here.

Given this bound,
the probability $P_{jump}(r>R_o)$ to have a jump distance $r$
larger than $R_o$ for large $R_o$ scales in the same
way as the probability
to have a backward avalanche of linear size $R$ larger than $R_o$,
or, alternatively, volume $S$ larger than $R_o^D$.
Therefore,
$P_{jump}(r>R_o) \sim P_b^{all}(S>R_o^D)=R_o^{-D(\tau_b^{all}-1)}$.
Substituting  the expression for $\tau_b^{all}$
from Eq. (\ref{e_all_b}) into this relation and differentiating with respect
to $R_o$, we get the final expression for the
exponent $\pi$,
\begin{equation}
\label{e_pi}
\pi=1+D(2-\tau) \qquad .
\end{equation}
This expression is in agreement with the result
$\pi=1+\gamma/\nu=1+\nu D (2-\tau)/ \nu=1+D(2-\tau)$ which was
derived for the Sneppen model using different methods in
\cite{ mppre}.
For invasion percolation, the relation between
$\pi$ and $\tau_b^{all}$ was first derived
by Roux and Guyon  \cite{ rouxguyon}.
Unfortunately they had an erroneous
expression for $\tau_b^{all}$.

Based on Eq. (\ref{e_pi}) and our numerical
values for $D$ and $\tau$, we predict $\pi = 3.26$
($\pi = 3.20$) in the one (two)
dimensional evolution model.
We have measured the exponent $\pi=3.23 \pm 0.02$
(Fig. \ref{f_pi_BS1}),  and
Jovanovic {\it et al}
\cite{ stanley} also measured $\pi= 3.1 \pm 0.2$ in
the one dimensional evolution model.  In the one (two) dimensional LIM,
$\pi=2.93 \pm 0.01$ (Fig. \ref{f_pi_LIM1}) (
$\pi=2.89 \pm 0.03$ (Fig. \ref{f_pi_LIM2})).
These results also agree
with Eq. (\ref{e_pi}) and the numerical values
we obtained for the exponents
$D$ and $\tau$.
For the one dimensional Sneppen model,  we predict
$\pi = 2.21$; Sneppen and Jensen measured $\pi=2.25 \pm 0.05$
\cite{ snepjensen}, while Tang and Leschhorn measured $\pi = 2.20 \pm 0.05$.
In two dimensions,  Falk {\it et al} \cite{ sneppentwod} measured
$\pi =2.2 \pm 0.2$ consistent with our prediction based on
their measured values
$\chi = 0.50 \pm 0.03$ and $\tau = 1.45 \pm 0.03$.
For IP-3 Furuberg {\it et al} measured $\pi=2.1 \pm 0.1$
\cite{ furuberg}, which is consistent with their
measured value $d_B=1.37$ and $D=1.82$ and Eqs.
(\ref{e_tau_b}) and (\ref{e_pi}).


\section{The Fractal Pattern of Activity}

For the SOC extremal  models discussed in this article, the dynamics
consists of a series of extremal events  following one after another.
At any given time step, $s$, there is one and only one lattice site where
activity occurs.
The extremal character of this dynamics lies in the fact
that this site is always selected by the {\it global} minimum
(or maximum) of some local driving force. The activity of the model
in the critical steady state is highly correlated. It forms an
anisotropic  fractal in $d+1$ ($d$ spatial and one temporal) dimensional
space \cite{ mpb94}.  An example of this fractal activity pattern for the
one dimensional evolution model was shown in Fig.
 \ref{one}.  We recall that
one characteristic exponent of this fractal is
its mass dimension $D$.
This dimension relates $S$, the total amount of activity
within a certain time interval to the spatial
extent, or range of activity, $R$,
through
\begin{equation}
 S \sim R^D \qquad .
\end{equation}
Due to the sequential character of activity,
$S$ is trivially
 equal to the number of time steps within a selected time interval.

We have already mentioned that in extremal models there exists a
purely geometrical fractal property of the pattern of activity.
The distribution of distances (Levy jumps) between subsequent minimal
sites,
$\vec{r}_{min}(s)$, obeys a power law $P_{jump}(r) \sim r^{-\pi}$.
This fractal property was described using backward avalanches
in Section IVE.

The pattern of activity has another feature where cuts
in different directions are fractals themselves.   Since the
pattern is anisotropic, cuts in different directions have different
mass dimensions.
The cuts in the spatial direction, at an arbitrary point in time, $s$,
are in fact trivial: there is only one active site at any given time.
Each cut has exactly one point of activity.
The fractal properties of cuts in the time direction, $s$,
at a given point in space,  are more interesting.
Looking at  Fig.\ \ref{one}, one can see that
the activity has a tendency to revisit sites.
At any given site, the activity is recurrent in time, and can be
considered to be a ``fractal renewal process'' \cite{ lowen}.
The collection of  return points on the one dimensional
time axis
forms  fractal with  dimension
$0 \leq \tilde d \leq 1$.
As in earlier sections, we assume that the  projection
of the activity pattern onto the original $d$ dimensional lattice
forms a dense, compact region of dimension $d$.  Thus we can consider
the activity pattern within a spatial region $R^d$, to be
composed of $R^d$
one dimensional fractal time lines.   In order to
encompass all of the activity, each time line has a
length $\sim R^D$.  The number
of points of the activity cluster which fall
on any given time line scales as $\tilde d$.  Here, we assume also that
there is only one dimension $\tilde d$ and no multifractal properties
of these points on each time line.
Thus the
quantity $R^dR^{D\tilde d}$, the total number of active sites
in all time lines, must scale in the same way as
the mass of the entire cluster, $S \sim R^D$.
This gives an exponent relation:
\begin{equation}
\label{e.tilde}
\tilde d=1-d/D \; .
\end{equation}
Note that the exponent $\tilde d$ is not defined for IP.

Complex spatio-temporal fractal patterns also can be observed in non-SOC
systems when their parameters are  fine tuned so that they become
critical.
These avalanche patterns again
are characterized
by the fractal dimensions of different cuts.
Exponent relations analogous
to  Eq. \ (\ref{e.tilde}) are obtained below.

For example, let us
consider
a large, finite directed percolation
cluster on a $d+1$ dimensional lattice.
A
part of such a cluster
is shown in Fig.\ \ref{twentyfour} for $d=1$.
  This cluster is asymmetric with respect to
the $t$ direction.  Recall that in DP all
sites are updated in parallel, so that $t \rightarrow t+1$
when all $n(t)$ sites have been advanced.
Self-similarity requires that the
duration $T$ scales with the spatial extent
in any one of the $d$ directions perpendicular to time,
$R$, as $T \sim R^z$ where $z$ is the usual dynamical exponent relating
space and time.
The total size of the cluster, $S$, scales with the
spatial extent as $S \sim R^D$, where $D$ is the avalanche dimension.
Unlike the previous sequential example,
we now have more than one active site at a given time step.
Thus, $T$ and $S$ represent different quantities and the corresponding
exponents $z$ and $D$, relating them to spatial size $R$ are
not equal to each other.
In order to compute $D$, usually the cluster is partitioned into $R^z$
equal time slices.  Each such slice contains $n_{act} \sim R^{d_{act}}$
points of the cluster.
This method of partitioning the cluster gives
$R^{D} \sim R^z R^{d_{act}}$, so the avalanche dimension
is given by $D=z+d_{act}$.
As in the previous example,
the avalanche cluster can be composed as $R^d$ one-dimensional
fractals, parallel to the time axis,
which each contain  $T^{\tilde d}$ parts of the cluster.
Consequently,
\begin{equation}
\label{e.tildedp}
 R^D \sim R^d R^{z \tilde d}\quad ; \quad
 \tilde d = {D-d \over z} \quad .
\end{equation}
Note that this relation contains Eq. \ (\ref{e.tilde}) as a limiting case
when $z=D$ and $d_{act}=0$ (one active site at any given time step).

Knowledge of the exponent $\tilde d$
enables us
to calculate two important distributions
characterizing return times of activity to a given point in space.
As a result, it also characterizes the power spectrum of local activity.
The first  return probability
distribution
$P_{FIRST}(\tilde t)$
 is the distribution of `hole' sizes, or intervals,
separating subsequent return points of
activity.
Here $\tilde t$ is the size of
the hole (either in parallel time $t$ or sequential time $s$).
This distribution is normalizable; $\int_0^{\infty}
P_{FIRST}(\tilde t) d\tilde t = 1$.
The average total number of return points, $n(T)$, in an interval of length
$T$ is given by the fractal dimension of return points as
$n(T) \sim T^{\tilde d}$.  It
can be related to the first return probability;
\begin{eqnarray}
n(T) = T - n(T)\int_1^T P_{FIRST}(\tilde t) \tilde t d\tilde t \;  , \;
{\rm where} \nonumber \\
 P_{FIRST}(\tilde t) \sim \tilde t^{-\tau_{FIRST}} \; {\rm for}
\; \tilde t \gg 1 \qquad .
\end{eqnarray}
If $\tau_{FIRST} \le 2$ then
the divergence at the upper limit must cancel the $T$ term, so that
$T \sim n(T)T^{2-\tau_{FIRST}}$.
 This leads to the scaling
relation,
\begin{equation}
\label{e.tau_f}
\tilde d = \tau_{FIRST} - 1 \qquad ,
\end{equation}
connecting the fractal dimension
of return points to the distribution of  hole sizes.

The second distribution, $P_{ALL}({\vec r},\tilde t)$ is the probability that
activity at position 0 at time 0  will be
at $\vec r$ at time $\tilde t$.
 This quantity does not obey the same normalization
condition as $P_{FIRST}$.  Instead
$\int P_{ALL}({\vec r},\tilde t) d{\vec r} =N$,
where $N$ is the average number of active sites.
$P_{ALL}(0, \tilde t)$ is
the probability for the activity at time $\tilde t$ to revisit
a site that was visited at time $0$.
Unlike $P_{FIRST}(\tilde t)$ it doesn't require that the return is
the first return of activity, so it is often
referred to as the distribution of {\it  all return} times.
Since $n(T)$ is simply the sum of all returns of activity to a particular
site up to time $T$, we have
\begin{equation}
 n(\tilde t+1) - n(\tilde t) =  P_{ALL}(0,\tilde t) \ .
\end{equation}
Assuming a power law form
 $ P_{ALL}(0,\tilde t) \sim \tilde t^{-\tau_{ALL}} \; {\rm for}
\; \tilde t\gg 1$, one gets for $\tau_{ALL}$
\begin{equation}
\tilde d =1-\tau_{ALL} \qquad .
\label{e.tau_a}
\end{equation}
Comparing equations (\ref{e.tau_f}) and (\ref{e.tau_a})
gives the general  relation
\begin{equation}
\label{e2}
\tau_{FIRST} + \tau_{ALL} = 2 \qquad
\quad {\rm for} \qquad  \tau_{FIRST} \leq 2 \qquad .
\end{equation}
connecting the ``lifetime'' exponents for
the first and all returns of activity.

Recently, Ito \cite{ ito}
has examined the International Seismological Center
data of California earthquakes and has found that the ``all'' and
``first'' return time distributions for earthquakes to return
to a given location are power laws over approximately
two decades with characteristic exponents that obey Eq. (\ref{e2});
i.e. $\tau_{FIRST} \approx 1.4$ and $\tau_{ALL} \approx 0.5$.
In addition, he found that the distribution of jumps between subsequent
hypocenters of earthquakes may also be a power law with an exponent
$\pi \approx 1.7$.  This is a remarkable demonstration of the generality
of our results.

\subsection{1/f Noise and Punctuated Equilibria}

Since $P_{ALL}(0,\tilde t)$ is
the autocorrelation function of the activity, the
power spectrum is simply
\begin{equation}
\label{e1/f}
 S(f) = \int_{-\infty}^{+\infty} P_{ALL}(0,\tilde t)e^{2\pi if\tilde t}d
\tilde t \sim
{1 \over f^{\tilde d}} \quad .
\end{equation}
The mathematical relationship between return times and the power spectrum
 was derived previously using different methods by
Lowen and Teich \cite{ lowen}.  Here we have shown that
$1/f$ type noise emerges naturally in both self-organized
and non-self-organized critical systems as a consequence of avalanche
dynamics.  As a result,
the  actual exponent $\tilde d$ that characterizes the noise is
determined by the dimension $D$ of the avalanches.
Eqs.
(\ref{e.tilde}), (\ref{e.tildedp}), (\ref{e1/f}) establish
 a formal connection
between $1/f$ noise and fractal scaling behavior, i.e. spatio-temporal
complexity,
 in both tuned and self-organized critical models.

Model dependent behavior occurs between the upper critical and lower
critical dimension.
In the mean field limit, or above the
upper critical dimension, the activity is barely able to return and
$\tau_{FIRST}=\tau_{ALL}=1$.  As a result, the power spectrum,
$S(f) \sim 1/f^0$, corresponds
to white noise.  On the other hand, at the lower critical dimension,
the activity becomes dense in time and $\tilde d \rightarrow 1$.  In this case,
the power
spectrum  $S(f) \sim 1/f$, with logarithmic corrections.

Eqs. (\ref{e.tilde} - \ref{e1/f}) were
checked by  numerical simulations.
We simulated bond DP on a square lattice
in 1+1 dimensions at $f=0.6445$ for $L=3000$. The data shown in Fig.
\ref{twentyfive}, $\tau_{FIRST} \simeq 1.86$ and
$\tau_{ALL} \simeq 0.14$,
are in  agreement with the theoretical prediction $\tau_{FIRST}
=
1.84 \pm 0.02$ and $\tau_{ALL} = 0.16 \pm 0.02$
 based on the exponents $D$ and $z$
  in Ref.  \cite{ oneddp} .  Also $S(f) \sim 1/f^{0.84}$ in 1+1 dimensions.
In $d=1$ we simulated the  BS branching process at
branching probability $f=0.667$ and averaged over $\approx 10^9$ mutations
 to obtain Fig. \ref{twentysix}.
 Our measured values are $\tau_{FIRST} = 1.58 \pm 0.02$
and $\tau_{ALL}= 0.42 \pm 0.02$, quite close to the predicted values 1.59
and 0.41, respectively.
Similar results were found for $d=2$, at branching probability $f=0.390$,
$\tau_{FIRST} \simeq 1.28$ and $\tau_{ALL} \simeq 0.70 $ to be compared with
1.31 and 0.69 from the formulae above.
 The predicted  power spectrum is $S(f) \sim 1/f^{0.59}$ in
$d=1$ and $S(f) \sim 1/f^{0.31}$ in $d=2$.
We measured the power spectrum $S(f) \sim 1/f^{0.58}$
for the one dimensional
evolution
 model, as shown in Fig. \ref{twentyseven}a, and $S(f) \sim 1/f^{0.31}$
for the two dimensional evolution model, as shown in Fig. \ref{twentyseven}b.

The evolution model was introduced in an attempt to model punctuated
equilibria in biological evolution. Fig. \ref{twentysixa} shows
the accumulated number of activations, or returns to a given
site,  for the one
 dimensional evolution model.   In terms of the model, these
accumulated returns
would roughly correspond to accumulated mutations in a given species.
 The resulting devil's staircase shows plateaus
of stasis interrupted by bursts of activity.
The plateaus have a power law distribution in sizes decaying
as $S^{-\tau_{FIRST}}$.  This staircase behavior is
qualitatively similar to the punctuated equilibrium
behavior observed for  the evolution of real species
\cite{ gould}.  This suggests that the punctuations for a single
species are correlated to the avalanches in the global ecology.


\section{Interface Depinning}

So far, the main emphasis in this article has been on
a class of extremal
SOC models.  In chapter IV,
we started  comparing these  extremal models
with their constant force parallel counterparts, where all currently active
sites are updated in parallel at each time step.
Here we discuss the relationship between  tuned
and SOC models of interface depinning in more detail.

The behavior of an interface driven in the presence of
quenched random pinning forces appears in a wide variety
of contexts.  These include, among others, fluid invasion in
a porous medium \cite{ porous},
the motion of magnetic domain walls \cite{ rfim}, flux
lines,
  or charge density waves \cite{ tangcdw},
\cite{ cdw},  \cite{ middletoncdw}
in the presence of quenched disorder.
The depinning transition in the charge density wave system has
previously been described in terms of avalanche dynamics in
sandpile models \cite{ tangcdw}, \cite{ middletoncdw}.
The notion of an external force can be easily incorporated into the rules
of the models we have discussed.  Applying an external
force $f_o$ means moving in parallel all sites having their
local "depinning" force $f_i<f_o$.
Below the threshold force $f_c$, the  interface
is pinned in one of many possible metastable states.
An external force greater than the
threshold causes the interface to move
with a finite average velocity which vanishes continuously
at a depinning transition.
The instantaneous velocity, proportional to the density
of active sites, fluctuates in time around its average value.

Instead of tuning the
force to the depinning threshold, the velocity may be tuned.
In this case, the force fluctuates in time, in such a way that the
instantaneous velocity is constant.
The depinning transition at constant velocity is
reached as the limit when the imposed velocity vanishes.
The instantaneous interface velocity in  discrete lattice models
is simply proportional to the current density of active sites:
$v=cn_{act}/L^d$, where $c$ is a constant of ${\cal O}(1)$.
We restore the rules of SOC extremal interface models
if we require that strictly one site is moving at any given time step
and this site is selected as most unstable in the whole system, i.e
the global minimum of $f_i$.
This corresponds to the interface velocity
$v_{ext} =c L^{-d}$.
Some of the critical properties of the model
driven at constant velocity
are the same as at constant force but others are different.
It is interesting to note that Wilkensen and Willemsen
\cite{ invasion} introduced  invasion percolation
 as a modification
of earlier models \cite{ lenormand} that were driven at constant force.
As they noted, the introduction
of extremal dynamics corresponds to the constant {\it velocity}
invasion process in the limit of vanishing velocity.

Theoretical studies of  interface depinning have
concentrated along two main directions.
One theoretical approach has been to
apply a functional renormalization group procedure
to a variety of
different depinning phenomena \cite{ middletoncdw},
\cite{ rg}.  This has yielded both
perturbative results and results that have been claimed to be
exact.  For example, for the LIM defined by  Eq. (\ref{e.force}),
it has been claimed \cite{ narayan.93}
that the roughness exponent $\chi = (4-d)/3$ exactly.
Another theoretical approach has been to consider simple lattice
models for these phenomena, which may, due to their simplicity,
be analytically tractable.

Several of the parallel lattice models defined in Section II describe
interface depinning at constant
force.
The LIM and TLB models presumably describe the depinning of magnetic domain
walls having purely linear force terms,
and other elastic interfaces having nonlinear force
terms, respectively \cite{ gallucio}.
Sneppen was the first to introduce a
self-organized critical extremal model
dedicated to interface
depinning; similarly an SOC, or
constant velocity, variant of the LIM model can be constructed
\cite{ mpb94}.
The general properties
of these SOC models have been analyzed in the preceding sections.
Our main results were the
derivation of two exact equations, a stationarity condition,
 and numerous exponent
relations.
These scaling relations are summarized in Table I.

We now discuss how our results may be generalized
to the tuned depinning models driven by constant
external force.
Here, we only analyze the behavior
as the depinning transition is approached from below.
In this limit, we will show that the tuned and SOC variants have
the same avalanche dimension $D$, the same avalanche distribution
exponent $\tau$, and the same roughness exponent $\chi$.
 However, the exponents that describe propagating
activity within avalanches are different.
In particular, the stationarity law $\eta=0$ does
not necessarily hold for the tuned models at constant force,
although it does hold for the
extremal SOC models.  This implies that the fractal dimension of
activity $d_f$ and the dynamical exponent $z$ can also be different in
these two cases.  These results are verified numerically.

We first concentrate on the case of the Sneppen and TLB models
which are constant velocity and constant force versions
of the same depinning phenomenon.
Let us recall the definition of the $f_o$- avalanche in the Sneppen model.
This avalanche
intervenes between subsequent punctuations of the barrier $f_o$ by
the signal $f_{min}(s)$, which is the extremal value of the pinning
force.  Geometrically, an $f_o$- avalanche takes the interface from
one critical ``blocking surface'' where all the random numbers are greater
than or equal to
$f_o$ to another $f_o$ blocking surface, as shown in
Fig. \ref{schematic}.  Tang and Leschhorn
\cite{ sneppentl} showed that in one dimension, these
blocking surfaces correspond to percolating paths on a
directed percolation cluster, formed by all sites with
$f_i<f_o$.  Given a collection
of random pinning forces $f({\vec x},h)$, and an initial interface
configuration identifying with a critical blocking
surface, the next blocking surface that is encountered
under Sneppen dynamics is the same blocking surface
that would be encountered in an equivalent parallel model, i.e.
the TLB model, driven with an applied force $f_o$.  The
order in which the sites between these two blocking surfaces
are invaded  is completely
different, though. In the Sneppen model, one always chooses the smallest
random pinning force, $f_{min}$.  In the TLB model, one advances all
sites where $f_i \leq f_o$ in parallel.  Nevertheless, the difference between
the initial interface configuration and the final configuration,
given by the two blocking surfaces, is the same
for the two models.
Thus the TLB model has the same threshold $f_c$, roughness exponent
$\chi$
of blocking surfaces, and avalanche distribution
$P(S,f_o)$.
This has been confirmed numerically
in Refs.  \cite{ sneppen}, \cite{ sneppentl},
\cite{ paralleltl}, \cite{ parallelstanley}, \cite{ sneppentwod}.

The roughness exponent $\chi$ characterizes the saturation width,
$w$, of the height fluctuations in a system of size $L$, so that
 $w^2 = <(h -<h>)^2> \sim L^{2\chi}$ \cite{ family}, \cite{ zhang_halpin}.
The roughness exponent $\chi$ and
the avalanche dimension $D$ are related via
\begin{equation}
D=d+\chi \qquad .
\label{e.D.chi}
\end{equation}
The logic behind this relation is:
 1) The set of sites advanced at least
once during the course of an avalanche is assumed to
form a compact object
having the same dimension $d$ as the interface substrate.  In
$d=1$ any connected set is also compact so this assumption is
obviously valid.  In higher dimensions, we are not aware of any proof
that the avalanches are compact.
2) It is assumed that there are no multifractal features of
the measure,
 defined as a total
number of updates during an avalanche,
 on the set of $n_{cov}$ sites.  Namely, there is only one scale
characterizing the number of times each site covered by the avalanche
is updated.
If both conditions 1) and 2) are satisfied, the avalanche
volume, which in extremal models is simply proportional to its
temporal duration $S$, can be written as $R^d R^{\chi}$, and
the relation (\ref{e.D.chi}) is satisfied.
On the other hand,
it is not completely inconceivable that in high enough dimensions
one of the assumptions can be wrong for the interface models.
In particular, this would occur in dimensions above the upper
critical dimension of the model, if an upper critical dimension
 exists.  Then Eq. (\ref{e.D.chi})
no longer holds, and $D$, the avalanche
dimension, can be  smaller than $d$, the dimension
of the substrate.  The results of the numerical simulations of the TLB
model reported in \cite{ amaral},
\cite{ parallelstanley} suggest that the relation
(\ref{e.D.chi}) may be valid in dimensions as high as $d=4$.

Taking into account these assumptions,
our results contained in section IV B, which were
derived for Sneppen model
directly apply to its parallel counterpart - the TLB model.

The critical dynamics of the interface as it propagates from
one blocking surface to another may be different in the
constant velocity (SOC) and constant force interface models.
This difference is not restricted to a trivial time redefinition,
where in extremal versions the temporal duration of an
avalanche is simply proportional to its volume, while in parallel
versions these quantities differ
 because all active sites are advanced in parallel.
What is more interesting is that even
the fractal dimension, $d_f$, of active sites
(sites with $f_i<f_o$) is different in these two cases.
Since the dynamics of the Sneppen model is stationary,
the dynamical exponent $\eta =0$.  There is no such stationarity condition
in the TLB model
at constant force; thus $\eta$ is not necessarily zero in that
model.  In fact, in one dimension Tang and Leschhorn \cite{ paralleltl}
have argued that $z=1$ exactly for TLB at constant
force, in contrast to $z \equiv D - d_f \simeq 1.21$
which is predicted as a corollary of the $\eta=0$ law for the
Sneppen model and which was measured in Fig. \ref{twenty}.
In fact, we measured
$z$ for the TLB model, as also shown in Fig. \ref{twenty}, and confirmed the
$z=1$ prediction and previous numerical results \cite{ paralleltl},
\cite{ parallelstanley}.
This  demonstrates that the dynamical critical exponents
in SOC and tuned critical phenomena can be different.

It is possible, though, that these differences may become smaller
or disappear entirely
for tuned and SOC versions
as the substrate
dimension increases. The numerical simulations of the TLB model in dimensions
up to $d=4$  \cite{ amaral}, \cite{ parallelstanley}
may possibly support this point of view.
One of the consequences of the $\eta=0$ law is that
the exponents they defined as
 $\tau_{surv}$ and $\delta$ would satisfy the relation
$\delta=d_f/z=\tau_{surv}-1$. This relation is obviously violated in
the one dimensional model where they
\cite{ amaral} report $\tau_{surv}^{(1+1)}=1.46(2)$ and
$\delta^{(1+1)}=0.60(3)$, which once again rules out $\eta=0$ for the
one dimensional TLB model.
Their results for higher dimensions
$\delta^{(2+1)}=1.14(6)$,
$\tau_{surv}^{(2+1)}=2.18(3)$,
$\delta^{(3+1)}=1.6(1)$,
$\tau_{surv}^{(3+1)}=2.54(5)$,
$\delta^{(4+1)}=1.9(2)$,
$\tau_{surv}^{(4+1)}=3.0(2)$,
however, seem to obey the $\eta=0$ relations within their numerical
uncertainties.  We do not know whether this apparent agreement in
dimension $d>1$ is simply a numerical coincidence.

A similar argument that in tuned and self-organized versions of
the model the exponents $\tau$ and $D$  are the same, while $\eta$
in general is different can be applied to the LIM model.
In this model, the advancement at any given site can never cause
a neighboring unstable site to become stable; the motion
at an active site can never destroy another active site.  This means
that one can interchange, arbitrarily, the order in which the unstable
sites move without changing the final
metastable configuration that is reached.  This property
is reminiscent of the Abelian properties of the  sandpile model
exploited by Dhar \cite{ dhar}.  This interchangeability means
that the critical force $f_c$, as well as the exponents $D$ and
$\tau$ describing the avalanche statistics, are the same for the
two versions.  Obviously, though, $\eta$ can be different because
the dynamics during an avalanche depends on the way unstable sites
 move.  In fact, Leschhorn measured $z= 1.42\pm 0.03$ in one dimension
and $z= 1.58 \pm 0.04$ in two dimension for the tuned
LIM at constant
force; whereas, based on our measured values of $D$ and $\tau$,
we predict quite different
values $z \simeq 1.94$ (one
dimension) and $z \simeq 2.04$ (two dimensions)
for the SOC or constant velocity variant.

For the interface models,
it is conventional to assume
the validity of the relation (\ref{e.D.chi}) and
to express the critical exponents in terms of
$\chi=D-d$, and $\nu$, the spatial correlation length
exponent along the substrate.
In this case, Eqs. (\ref{e.int.tau}-
\ref{e.int.gamma}) can be rewritten in terms of
$\nu$ and $\chi$ as
\begin{equation}
\tau=1+{d-1/\nu \over d+\chi}
\label{e.tauINT}
\end{equation}
\begin{equation}
\gamma=1+\chi\nu \qquad .
\label{e.gammaINT}
\end{equation}
In the one-dimensional Sneppen model, $\nu$ and $\chi$
have been derived from the fractal properties of the
1+1-dimensional directed percolation
cluster \cite{ sneppentl}, \cite{ mppre}.
Namely $\chi=\nu_{\perp}^{DP}/\nu_{\parallel}^{DP}$
and $\nu=\nu_{\parallel}^{DP}$, where $\nu_{\perp}^{DP}$ and
$\nu_{\parallel}^{DP}$ are parallel and perpendicular correlation
length exponents in 1+1-dimensional directed percolation.
For the one dimensional Sneppen model, these expressions differ
from the predictions of Olami, Procaccia, and Zaitek that
$\tau = 2/(2+\chi)$ and $\gamma=2$ \cite{ olami.94}.
The avalanche dimension $D$ was measured for the self-organized
LIM as shown in
Fig. \ref{fourteena}.
In $d=1$, $D = 2.23 \pm 0.03$, and
in $d=2$, $D = 2.725 \pm 0.02$.
For the LIM in both one and two dimensions, $D$ was measured
by computing, for each backward avalanche, the duration $S$ of
the backward avalanche, and the squared distance $R^2$ from the
site starting  the backward avalanche to the site ending it.
Substituting these measured values
into the relation $\chi= D -d$, gives  $\chi \simeq
1.23$ in $d=1$ and $\chi \simeq 0.72$ in $d=2$.  These values for
$\chi$ are in agreement with numerical simulations by Leschhorn
\cite{ leschhorn},
but higher than the prediction $\chi = (4-d)/3$
from functional renormalization
group (RG) calculations \cite{ narayan.93} in both one and
two dimensions.

One might wonder if the one dimensional LIM describes
the behavior of any real physical system. Our measured value of the
interface
roughness $\chi=1.23 \pm 0.02$ \cite{ mpb94}
derived from the avalanche fractal
dimension $D=1+\chi$, as well as  the reported values
$\chi=1.25 \pm 0.01$ \cite{ leschhorn} and $\chi=1.2 \pm 0.1$
\cite{ roux}, contradicts the usual condition that the "self-affine"
interface looks flat when viewed at a sufficiently large length scale.
It may be more appropriate to call the interfaces with $\chi>1$ "super-rough"
\cite{ zhang_halpin}, \cite{ superrough}.
As was suggested in \cite{ roux}, one possible physical realization
of the super-rough
interface is  fluid invasion in a porous medium.  Martys, Robbins,
and Cieplak \cite{ martys} introduced an explicit model for this process.
They have shown that depending on the wetting
properties of the invading fluid in 1+1 dimensions one observes two
distinct universality classes of the constant pressure, i.e. force, interface
depinning transition.  One was identified with the constant pressure
version of invasion percolation \cite{ lenormand},
having an extremely curved interface with
overhangs at all length scales. The interface from
the other universality class has overhangs
only at small length scales, and Roux and
Hansen have argued that the 1+1 LIM shares the same universality class
\cite{ roux}.
Our numerical results support this conjecture.
We have measured $\tau=1.13 \pm 0.03$ in very good agreement with
the measured value
$\tau=1.125 \pm 0.025$ of \cite{ martys}.
Scaling relations analogous to Eqs.
(7, 31, 32)
for the non SOC model
were derived in \cite{ martys} based on the assumption that
the avalanches in the model are isotropic; specifically they
assumed $D=d+1$, which is different than our result
$D=d+\chi$.
Nevertheless, these authors considered the
interface itself   to be self-affine ($\chi \neq 1$),
and not isotropic.
Within numerical
uncertainty, it appears to
us that  their results are also consistent with
anisotropic super-rough avalanches having the
same fractal properties
as the interface itself ($\chi=D-d$).  For example,
inserting their numerical value $\nu=1.30 \pm 0.05$
and our numerical value
$D=2.23$  into Eq. (\ref{e.tauINT}) one gets $\tau=1.105$,
which is also
consistent with the measured values cited above.
The question of the shape of the avalanches in their model
could probably be resolved by  direct measurements of the avalanche
volume $S$ vs. its spatial extent $R$.

Similiar considerations may apply to self-affine interfaces
in the 2+1-dimensional Random Field Ising Model studied by Ji and
Robbins \cite{ rfim}.  They  measured $\tau=1.28 \pm 0.05$ and
$\nu=0.75 \pm 0.05$ in agreement with our numerical result $\tau=1.29
\pm 0.02$ for the 2+1-dimensional LIM and Eq. (\ref{e.tauINT}) using
$D=2 + \chi=2.72 \pm 0.02$.
Again, they assumed the avalanches to be isotropic in
the Random Field Ising Model, while
the avalanches in the LIM are anisotropic and self-affine.

\section{Conclusions}

We have presented a
comprehensive theory for avalanche dynamics
in evolution, growth, and depinning models.  These models are
defined in Section II and represent different universality
classes.  We have shown that
avalanche dynamics leads to spatio-temporal complexity, and
emerges as a result of extremal dynamics
in driven systems.  Spatio-temporal complexity is manifested in
the formation of fractal structures, the appearance
of $1/f$ type noise, diffusion with anomalous Hurst exponents, Levy
flights, and punctuated equilibrium behavior.
These phenomena can all be related to the
same underlying avalanche dynamics.

We present two  exact equations for
these phenomena. (1)  The approach to the critical attractor is governed
by a ``gap'' equation for the divergence of avalanche sizes.  (2) The
 average number of sites covered by the avalanche can be related to the
average size of avalanches.  If there is a power law divergence
for the average avalanche size, then
the number of sites covered by an avalanche diverges with
exponent -1.
In addition,
the conservation of activity in the stationary state manifests itself
through the fundamental relation $\eta = 0$.  It follows
that many of the
critical exponents in a class of SOC extremal models
can be derived from two basic exponents.

These exponent
relations are summarized in Table I.
Depending on the model it may be more convenient
to use one or another basic set
of two exponents.
In Table I we write our exponent relations in terms
of two such basic sets: ($D$, $\nu$), or ($D$, $\tau$).
The scaling relations are defined for the Bak-Sneppen evolution model,
the Sneppen model for SOC interface depinning,
the Zaitsev model flux creep model, an SOC ``linear'' interface
model, and invasion percolation.
The horizontal lines separate results from different sections of
this article.

Our results from numerical simulations are summarized in Table II.
This Table contains the expressions for the exponents of the
one and two dimensional
Bak-Sneppen model, one and two
 dimensional LIM, and the one dimensional Sneppen model
based on {\it direct} numerical simulations.  The
overall consistency between
 the exponent relations from Table I with the numerical results in Table II
demonstrates the validity of our new scaling relations.
Most of the numerical results from Table II
were illustrated in figures throughout the
text.

Some of these new scaling relations
show that
$1/f$ noise and spatial fractal behavior can be unified, and
have a natural explanation in terms of avalanche dynamics in both
SOC and non-SOC systems.
The pattern of avalanches in the  critical
state
can be described as a fractal in $d$ spatial plus one temporal dimension,
with mass dimension $D$.
Temporal behavior, such as $1/f$ noise in the power spectrum,
and spatial long-range correlations
can be formally related as different cuts in the
same underlying fractal.
In the stationary state, time reversal symmetry is
broken, so that forward and backward avalanches have different statistics.
We derive a scaling relation for the Levy distribution of jumps in the
SOC extremal models
models.  Finally, we point out the similarities and differences between
interface depinning at constant velocity (SOC)  and constant force.

This work was supported by the
U.S. Department of Energy
Division of Materials Science,  under contract DE-AC02-76CH00016. MP
thanks the U.S. Department of Energy Distinguished Postdoctoral Research
Program for financial support.  The authors thank K. Sneppen
 for useful conversations, and B. Drossel for helpful comments on the
manuscript.


\section{Appendix A: Invasion Percolation}
Invasion percolation  is very sensitive to the
definition of the boundary of the invaded cluster. In  one
version of the model, the boundary consists of any site having
at least one nearest neighbor in the invaded cluster.
The boundary  defined by this rule includes
sites on the external perimeter of the invaded cluster as well as
  sites on the perimeter of numerous "lakes"
in the interior of the cluster.
The invaded cluster in this model
from time to time exactly identifies itself with the infinite cluster of
ordinary percolation.
At these instances, all of the random numbers on the boundary
are uniformly distributed above the ``gap'', $f_c$, where $f_c$ is the
critical density of ordinary percolation.
In two dimensions, the whole boundary of the cluster of ordinary percolation
has a fractal dimension, $d_B$, equal to the fractal dimension $D$ of the
cluster itself, and $D=d_B=91/48 \simeq 1.89$
\cite{ feder}.
We will refer to the invasion percolation model with the boundary
defined in this manner as IP-1.

An important physical realization of invasion percolation is
the displacement of one fluid by another in a porous medium.
In this case, once a region of non-invaded fluid is completely surrounded
by the invading fluid, no further invasion can take place on this part
of the boundary due to incompressibility.  This event, known as
self-trapping, can be taken into account in the invasion percolation model
by changing the definition of the active boundary.
With self-trapping,
only the sites touching the {\it external perimeter} of the
invaded region comprise the
active boundary.  It is believed that self-trapping does not change
the fractal dimension of the invaded cluster in two dimensions
\cite{ feder}.
In ordinary two dimensional percolation,
it is known that the external perimeter of the cluster has a smaller fractal
dimension than the cluster itself, i.e. $d_B < D$.
Depending on the details of
the precise definition of the external perimeter, one gets
$d_B=7/4=1.75$ \cite{ saleur},
or $d_B=4/3 \simeq 1.33$ \cite{ grossman}.
  These different definitions thus give
different variants of the invasion percolation model, which we will refer
to as IP-2, and IP-3, respectively.

In Section III we used the result by Roux and Guyon \cite{ rouxguyon} for the
scaling of the boundary of
the  growing IP cluster with its volume.  For the sake of
completeness, we reproduce their arguments.  During the
transient, the invaded cluster
can be characterized by a growing correlation length, $\xi$.
Since the growth starts  from a $d$-dimensional base
 of the $d+1$-dimensional hypercube,
the natural scaling form for the invaded volume $s$  is
\begin{equation}
\label{e_vol_ip}
 s \sim (L/\xi)^d\xi^D \ ,
\end{equation}
and for the  boundary of invaded cluster, $b(s)$,
\begin{equation}
 b(s) \sim (L/\xi)^d\xi^{d_B} \quad .
\end{equation}
Combining these two equations we get:
\begin{equation}
 b(s)/L^d=(s/L^d)^{d_B-d \over D-d} \qquad ,
\end{equation}
and therefore $g={d_B-d \over D-d}$.

It is interesting to note that the simplest assumption that the effective
distance from the critical point scales as  $f_c-G(s)\sim \xi^{-1/\nu} \sim
(s/L^d)^{-1/\nu(D-d)}$ agrees with the gap equation  after
we substitute into it the expression for $g$ and for $\gamma$.
This once more confirms the overall consistency of our scaling relations.



\def\fig#1\par{\begin{figure}\caption{#1}\end{figure}}

\fig\label{one}%
 The space-time fractal activity pattern for the one dimensional
evolution model. Time is measured as the number of update steps, $S$.%

\fig\label{two}%
 Snapshot
of the stationary state for the evolution model
 in a finite one dimensional system.
Except for a localized region, the avalanche, where there
are small random numbers, all the random numbers in the system have
values
above the self-organized threshold $f_c = 0.66702$.

\fig\label{schematic}%
 Schematic picture of an avalanche separating two interface configurations
in the Sneppen model. The size, $S$, of the avalanche corresponds to
the shaded area.

\fig\label{three}%
Value of the extremal signal $f_{min}$  as a function of $s$ during
the transient in a small  (L=20) one dimensional evolution model.
 The full line shows
the gap, $G(s)$, as a step-wise
 increasing function of $s$.   The gap is an envelope function that tracks
the peaks in $f_{min}$.

\fig\label{five}%
 Illustration of the hierarchical structure
of the $f_o$- avalanches. The big avalanche is subdivided
into smaller ones as the auxiliary parameter $f_o$ is lowered.
The lines span $f_o$- avalanches for $f_o=0.63, 0.59$, and $0.54$,
 respectively.

\fig\label{six}%
Distribution of avalanches from simulation of the BS branching process in
a) one and b) two dimensions. The asymptotic slope of the log-log plot
gives $\tau = 1.07 \pm 0.01$ in 1d and $\tau = 1.245 \pm 0.01$ in 2d.

\fig\label{seven}
 One realization of the extremal signal
$f_{min}$  as a function of $s$ in the stationary state.
If the auxiliary parameter $f_o$ is raised by an infinitesimally
small amount $df_o$, the breaking points that had $f_{min}(s)$
between $f_o$ and $f_o+df_o$ (filled circles)
 will no longer stop the $f_o+ df_o$-  avalanches,
and the average avalanche size will increase.

\fig\label{eight}
 The gamma plot
 $(1-f_o)/<n_{cov}>$ vs. $f_o$ for the two dimensional evolution model.
The intersection with the horizontal axis gives $f_c$, and
the slope gives $1/\gamma$. We find  $f_c = 0.328855 \pm 0.000004$
and  $\gamma = 1.70 \pm 0.01$.

\fig\label{nine}
 As Fig. 8 for the 1d evolution model.  We find  $f_c = 0.66702 \pm 0.00003$
and  $\gamma = 2.70 \pm 0.01$. \hfill

\fig\label{ten}
 The plot of $(1-f_o)/ <n_{cov}>_{f_o}$
as a function of $f_o$ for the one dimensional Sneppen model.
The slope of the straight line is $1/\gamma=0.470$ and the crossing
point with $f_o$-axis correspond to $f_c=0.465$.

\fig\label{eleven}
 The average size of avalanches,
$<S>$, vs. $(f_c - f)$ for the 2d evolution model.
The asymptotic slopes yields $\gamma = 1.69 \pm 0.03$.

\fig\label{twelve}
 The average size of avalanches,
$<S>$, vs. $(f_c - f)$ for the 1d evolution model.
The asymptotic slopes yields $\gamma = 2.71\pm 0.02$.

\fig\label{fourteen}
  a)
$R$ vs. $S$ for the 2d evolution model (averaged over 107000 avalanches).
b) Derivative of this curve.  The asymptotic value corresponds to
$1/D$ and gives
$D = 2.92 \pm 0.02$.

\fig\label{fifteen}
 $R$ vs. $S$ for the 1d evolution model involving $ 3\times 10^{11}$ updates.
{}From this data we extract
$D = 2.43 \pm 0.01$. \hfill

\fig\label{fourteena}
 $S$ vs. $R$ from backward avalanches in 1d (open circles) and
2d (filled circles) LIM.
In 1d, the slope is $D = 2.23 \pm 0.03$ from simulations of
$5 \times 10^7$ time steps in the stationary state of
 a system of size $L=3000$.
In 2d,
the slope is $D = 2.725 \pm 0.02$ from simulations of
$10^9$ time steps in the stationary state of
a system of size $300 \times 300$.

\fig\label{sixteen}%
Typical n(s) for the two dimensional evolution model.

\fig\label{seventeen}%
Average number of active sites $<n(S)>$ vs $S$ (lower curve)
and average activity of surviving avalanches, $n_{SURV}(S)$ (upper curve),
for the two dimensional evolution model at $f_{c} = 0.3289$. The slopes
yield $d_{s} = 0.25 \pm 0.005$ and $\eta =0.0 \pm0.002$ respectively.

\fig\label{eighteen}%
Average number of active sites $<n(S)>$
and average activity of surviving avalanches, $n_{SURV}(s)$,
for the one dimensional evolution model at $f_{c} = 0.6670$. The slopes
yield $d_{s} = 0.11 \pm 0.02$ and $\eta =0.01 \pm 0.02$.

\fig\label{twenty}%
Parallel time, $t$, vs. spatial extension, $r$,  for the one dimensional
Sneppen model (open dots) and the one dimensional
TLB model (filled dots).  The slopes gives $z=1.19 \pm 0.05$ for the
Sneppen model and $z=1.00 \pm 0.05$ for the TLB model.

\fig\label{twentyone}%
 Parallel time, $t$, vs. spatial extension, $r$, for a the two dimensional
(small filled dots)  and  one dimensional evolution (large open dots)
 models.
The asymptotic slopes are $z = 2.17 \pm 0.02$  and
$z = 2.10 \pm 0.05$, respectively.

\fig\label{twentytwo}%
$<n(S)>$  and $n_{SURV}(S)$ for the one dimensional Zaitsev model
with $L=10^4$.
$d_s = 0.15 \pm 0.03$.  The asymptotic value of
the slope of $<n(S)>$ is
$\eta = 0.0 \pm 0.006$.

\fig\label{fback}
 The hierarchical structure of backward and forward
avalanches. The avalanches that started inside a
larger  parental
avalanche are completely contained within it.

\fig\label{f_ba_snep1}
The overall distribution of backward avalanches
in the 1d Sneppen model. The slope of the curve
gives $\tau_{b}^{all}=3-\tau=1.745 \pm 0.02$.
We simulated $10^7$ time steps in a system of
size $L=1000$.

\fig\label{f_ba_LIM1}
The overall distribution of backward avalanches
in 1d (open circles) and 2d (closed circles) LIM.
In 1d, the slope
 of the curve
corresponds to $\tau_{b}^{all}=3-\tau=1.87$ from simulations of
$5 \times 10^7$ time steps
in a system of size $L=3000$.
In 2d, the slope of the curve
corresponds to $\tau_{b}^{all}=3-\tau=1.71$ from simulations of
$10^9$ time steps
in a system of size $300 \times 300$.

\fig\label{sneppenavalanche}
 Distribution of avalanches from simulations of the 1d
Sneppen model for $L=2\times 10^5$ and $f_c=0.4614$.
 The asymptotic slope of the log-log plot
gives $\tau = 1.26 \pm 0.01$.

\fig\label{f_pi_BS1}%
 The distribution of jumps between subsequent minimal sites
in the 1d evolution model. The slope of the straight line
corresponds to $\pi=3.23$.
We simulated $5 \times 10^7$ time steps
in the system of size $L=3000$.

\fig\label{f_pi_LIM1}%
 The distribution of jumps between subsequent minimal sites
in the 1d LIM. The slope of the straight line
is $\pi=2.93$.
We simulated $5 \times 10^7$ time steps
in the system of size $L=3000$.

\fig\label{f_pi_LIM2}%
The distribution of jumps between subsequent minimal sites
in the 2d LIM. The slope of the straight line
corresponds to $\pi=2.89$.
We simulated $10^9$ time steps
in a system of size $300 \times 300$.

\fig\label{twentyfour}
 The pattern of active sites in 1+1-dimensional bond directed percolation.

\fig\label{twentyfive}
 Distribution of return times for bond DP on a square lattice for
$f=0.6445$ and $L=3000$.
The asymptotic slopes give $\tau_{FIRST} \simeq 1.86$ and
$\tau_{ALL} \simeq 0.14$.

\fig\label{twentysix}
Return times for the a) 1d BS branching process at
branching probability $f=0.667$ averaged over $\approx 10^9$ mutations.
The corresponding exponents are $\tau_{FIRST} \simeq 1.58$
and $\tau_{ALL} \simeq 0.42$.
b) 2d BS branching process at
$f=0.3289$ averaged over $\approx 10^9$ mutations.
$\tau_{FIRST} \simeq1.28$
and $\tau_{ALL} \simeq 0.70$.

\fig\label{twentyseven}%
 Power spectrum for a) the one and b) the two dimensional
evolution model.

\fig\label{twentysixa}
 Punctuated equilibria. The curve shows the accumulated number of times
a specific site is visited, or is the minimal site,
 in the one dimensional evolution model.
This would roughly correspond to the accumulated
number of mutations in a particular species.  The pattern of change
is step-wise rather than being smooth.

\vfill\eject

\widetext
\tightenlines
\begin{table}
\caption{Exponent relations for the extremal models in $d$ dimensions.
For invasion percolation, $d$ should be replaced with the fractal
dimension of the active boundary $d_B$ except in the scaling
relation for $\rho$ (see text). The exponents $\tau_{ALL}$,
$\tau_{FIRST}$, and $\tilde{d}$ are not defined for invasion
percolation. In the
interface models $D$ is related to the interface roughness $\chi$ via
$D=d+\chi$.}
\label{ scalingrelations}

\begin{tabular}{cccc}

  Exponent & Physical Property & $(\nu$ and $D$) & ($\tau$ and $D$) \\
\tableline
  $D$ & Avalanche Dimension & basic exponent & basic exponent \\
  $\nu$ & Correlation Length & basic exponent&${1 \over d-D(\tau-1)}$ \\
  $\tau$ & Avalanche Distribution & $1+{d-1/\nu \over D} $ & basic exponent \\
  $\gamma$ & Average Avalanche Size&  $1+\nu(D-d) $ & ${2-\tau \over
1+d/D-\tau}$ \\
  $\sigma$ & Avalanche Cutoff & $1/\nu D$ & $1+{d \over D} -\tau$ \\
   $\rho$ & Relaxation to Attractor & ${1 \over \nu(D-d)}$ &
${ 1+{d \over D} -\tau \over 1 -d/D}$ \\
\tableline
  $\eta$ & Average Growth of Activity & 0 &0 \\
  $d_f$ & Dimension of Active Sites & $d-{1 \over \nu} $ & $D(\tau-1)$ \\
  $z$ &  Dynamical Exponent & $D-d+{1 \over \nu} $ & $D(2-\tau)$ \\
\tableline
  $\pi$ & Jump of Minimal Site & $1+D-d+{1 \over \nu}$ & $1+D(2-\tau)$ \\
  $\tau_f^{all}$ & All Forward Avalanches & 2 & 2 \\
  $\tau_b^{all}$ & All Backward Avalanches & $2-{d-1/\nu \over D} $ &
$3-\tau$\\
\tableline
  $\tau_{FIRST}$& Punctuated Equilibrium &$2-{d \over D}$ & $2-{d \over D}$ \\
  $\tau_{ALL} $ & All Returns &$ {d \over D}$ & $ {d \over D}$ \\
  $\tilde{d}$ & $1/f$ Noise & $1-{d \over D}$& $1-{d \over D}$\\
\end{tabular}
\end{table}

\begin{table}
\caption{Critical exponents measured and illustrated in figures
throughout the text.  All values were determined independently.
($\cdots $) indicates uncertainty in the last digit. Within
 these uncertainties, all exponents are consistent with the exponent
 relations from Table I.}
\label{ numerics}
\begin{tabular}{cccccc}
Exponent & \multicolumn{2} {c} {Bak-Sneppen model} & \multicolumn{2} {c}
{Linear Interface model} &
Sneppen model \\
& 1 dimension & 2 dimensions & 1 dimension & 2 dimensions & 1 dimension\\
& $f_c=0.66702(3)$ & $f_c=0.328855(4)$ &&& $f_c=0.4614(4)$\\
\tableline
D&2.43(1) &2.92(2) &2.23(3) &2.725(20) &\\
$\tau$&1.07(1) &1.245(10) &1.13(2) &1.29(2) &1.26(1) \\
$\gamma$&2.70(1) &1.70(1) &&&2.13(20)\\
$\pi$ &3.23(2) & &2.93(3) &2.89(3) & \\
$\eta$ &0.01(2) &0.000(2)&0.000(6) \tablenotemark[2]
&&0.03(5)\\
$d_s$ & 0.11(2) &0.250(5)&0.15(3)\tablenotemark[2] &&
0.26(2)\\
$z$ &2.10(5) &2.17(2)&&&1.19(5) \tablenotemark[3] \\
$\tau_{FIRST}$ &1.58(2)&1.28(3) &&&\\
$\tau_{ALL}$ &0.42(2)&0.70(3)&&&\\
$\tilde{d}$ &0.58(3) \tablenotemark[4]&0.31(3)
\tablenotemark[4]&&&\\
\end{tabular}
\tablenotetext[2] {Measured for 1d Zaitsev  model, presumably from
the same universality class \cite{ roux}.}
\tablenotetext[3] {This exponent for the TLB model (parallel analog of
Sneppen model) was measured to be $1.00 \pm 0.05$.}
\tablenotetext[4] {This exponent was measured from the power spectrum of
activity.}

\end{table}


\begin{references}
\bibitem{ commenta}
The exponent relations derived in this chapter can be generalized
to the case where the set of covered sites
 form a fractal of dimension $d_{cov}<d$ rather than being
compact.  This is accomplished by replacing $d$ with
$d_{cov}$ in all formulas in this section.
\bibitem{ remark}
Note that our definition of $\eta$ differs from the usual
definition of $\eta$, which is defined with respect
to parallel time, as in Ref. \cite{ Grassb79}, and which we refer to
as $\eta_t$.  The quantity which we call
$\eta$ is equal to $z\eta_t/D$.
\end{references}
\end{document}